\documentclass[aps,pre,preprint,groupedaddress]{revtex4}
\usepackage{graphicx} 
\usepackage{dcolumn} 
\usepackage{bm}
\newcommand{\ignore}[1]{} 
\usepackage{epsfig} 
\usepackage{color}
\usepackage{amsmath,amssymb}

\begin{document}

\title{A multi-state voter model with imperfect copying}

\author{Federico Vazquez}
\email[]{fede.vazmin@gmail.com}

\affiliation{Instituto de C\'{a}lculo, FCEN, Universidad de Buenos Aires and CONICET, Buenos Aires, Argentina}

\author{Ernesto S. Loscar}
\author{Gabriel Baglietto}

\affiliation{Instituto de F\'{i}sica de L\'{i}quidos y Sistemas Biol\'ogicos (IFLYSIB), UNLP, CCT La Plata-CONICET, Calle 59 no.~789, B1900BTE La  Plata, Argentina}

\date{\today}

\begin{abstract}
The voter model with multiple states has found applications in areas as diverse as population genetics, opinion formation, species competition and language dynamics, among others.  In a single step of the dynamics, an individual chosen at random copies the state of a random neighbor in the population.  In this basic formulation it is assumed that the copying is perfect, and thus an exact copy of an individual is generated at each time step.  Here we introduce and study a variant of the multi-state voter model in mean-field that incorporates a degree of imperfection or error in the copying process, which leaves the states of the two interacting individuals similar but not exactly equal.  This dynamics can also be interpreted as a perfect copying with the addition of noise; a minimalistic model for flocking.  We found that the ordering properties of this multi-state noisy voter model, measured by a parameter $\psi$ in $[0,1]$, depend on the amplitude $\eta$ of the copying error or noise and the population size $N$.  In the case of perfect copying $\eta=0$ the system reaches an absorbing configuration with complete order ($\psi=1$) for all values of $N$.  However, for any degree of imperfection $\eta>0$, we show that the average value of $\psi$ at the stationary state decreases with $N$ as $\langle \psi \rangle \simeq 6/(\pi^2 \eta^2 N)$ for $\eta \ll 1$ and $\eta^2 N \gtrsim 1$, and thus the system becomes totally disordered in the thermodynamic limit $N \to \infty$.  We also show that $\langle \psi \rangle \simeq 1- 1.64 \, \eta^2 N$ in the vanishing small error limit $\eta \to 0$, which implies that complete order is never achieved for $\eta >0$.  These results are supported by Monte Carlo simulations of the model, which allow to study other scenarios as well.
\end{abstract}

\maketitle

\section{Introduction}
\label{Intro}

Stochastic models of evolution have been successfully applied in various disciplines to study the dynamics of systems composed by many interacting entities such as genes in population genetics, animal or plant species in ecology, and people in linguistics and sociology, among others (see \cite{Blythe-2007} for a statistical physics review).  The most basic --neutral-- version of each of these models implements some type of copying mechanism by which an entity is removed and replaced by an exact copy of another entity in the population.  For instance, in a single step of the Moran model \cite{Moran-1958} for genetic drift (similar to the Wright-Fisher model \cite{Fisher-1930,Wright-1931}) a gene is chosen at random to die and replaced by a new gene that is a replica of another gene in the population, its ``parent'', also chosen at random.  Similarly, neutral models for the evolution of species in ecology consider that when a tree dies is replaced by an ``offspring'' of a randomly chosen tree in the forest \cite{Hubbell-2001}.  A theory analogous to that of population genetics was presented in \cite{Baxter-2006} to explore the dynamics of language change in the context of linguistic variables, such as vowel sound or grammar.  The copying mechanism is also used in the voter model for opinion formation \cite{Clifford-1973,Holley-1975}, where each individual adopts the opinion of one of its neighbors in the population.  More recently, this type of social imitation rule was introduced to study the flocking dynamics of a large group of animals \cite{Baglietto-2018}, for instance birds, where each bird aligns its flying direction with that of a nearby random bird.  In the case of all-to-all interactions, this flocking voter model is equivalent to the well known multi-state voter model (MSVM) \cite{Starnini-2012,Pickering-2016} for opinion dynamics, where the moving direction of a bird is associated to its opinion or decision.  The MSVM considers a population composed by a fixed number of agents (voters) subject to pairwise interactions, where each voter can hold one of $S$ possible states that represent different opinions or positions on a given issue.  In a single step of the dynamics, a voter chosen at random updates its state by copying the state of another agent randomly chosen in the population.  The MSVM assumes that the copying process is perfect, in the sense that once an agent copies the state of its partner these two agents are considered to be indistinguishable.  However, in a real life situation one would expect some degree of inaccuracy in the copying process that translates into an \emph{imperfect copying}.  For instance, a person can try to adopt the exact opinion of a partner on a given opinion spectrum, but the imitation may not be perfect and the agent ends up taking an opinion very similar but not equal to that of its partner.  The source of error in the copying process may also come from the fact that the perception of a person on its partner's opinion may not be completely accurate.

The imperfect social imitation was recently modeled by adding an external noise in the original voter model, to study the outcome of electoral processes \cite{Fernandez-2014}.  The original noisy $2$-state voter model assumes that, besides the copying dynamics, voters can randomly switch state.  This variant of the model was introduced independently some years ago to study phenomena as diverse as heterogeneous catalytic chemical reactions \cite{Fichthorn-1989,Considine-1989}, herding behavior in financial markets \cite{Kirman-1993} and species competition in probability theory \cite{Granovsky-1995}.  The study of the effects of noise in the voter model has lately gained attention in the physics literature.  Recently, the $2$-state noisy voter model has been explored in complex networks \cite{Carro-2016,Peralta-2018,Peralta-2018-1}, and its dynamics has also been investigated under the presence of zealots \cite{Khalil-2018} and the influence of contrarians \cite{Khalil-2019}.  In \cite{Diakonova-2015} the authors have found that noise changes the properties of the fragmentation transition observed in a coevolving version of the voter model \cite{Vazquez-2008-1,Demirel-2014} and the MSVM on complex networks \cite{Bohme-2012}.

A mechanism of imperfect imitation was implemented in \cite{Roca-2009} within a game theory model to study the dynamics of cooperation, where the process of adopting the strategy of a neighboring player combines two different imitation dynamics, the unconditional imitation and the replicator rule \cite{Szabo-2007}.  They found that cooperation is enhanced when the probability of choosing the replicator rule (the perturbation) adopts intermediate values.  In the context of flocking dynamics, it is reasonable to assume that birds make an error when trying to align with a close by bird, which is modeled by adding a small perturbation (noise) to the alignment process as in Vicsek-type models \cite{Vicsek-1995,Baglietto-2009}.  It is observed that the noise amplitude induces a transition from a --nematically-- ordered phase for low noise to a disordered phase for high noise.   

In this article we study a system of interacting particles subject to a multi-state voter dynamics with imperfect copying on a complete graph (all-to-all interactions).  For concreteness we use the language of flocking, where the states of particles represent a finite set of angular directions equally spaced in the interval $[0,2 \pi)$.  In a single iteration step of the dynamics, a particle chosen at random adopts a state that is contained in an interval centered in the state of another randomly chosen particle.  Thus, the level of the imperfection in the imitation process is given by the length of the error interval, which is a variable of the model.  We note that the update of a particle's state can also be thought as a two-step process where, in a first step, the particle copies the state of another particle and then, in a second step, its state is perturbed within an interval (spontaneous transitions between states).  Although the MSVM with imperfect copying studied here falls in the category of the noisy $2$-state voter models mentioned above, it exhibits some crucial differences with them.  That is, the multiplicity of states in the MSVM allows for different types of spontaneous transitions between states, which go beyond the stochastic transition in binary models.  Specifically, we consider a system where states are ordered (a discrete set of angles ordered from $0$ to $2\pi$) and noise-induced transitions are allowed only between neighboring states, and not between any two states as in most genetic models with mutations \cite{Blythe-2007}.  

We investigate the ordering dynamics of the system by numerical simulations and analytical techniques and found that the imperfection in the copying mechanism changes completely the ordering properties.  When imitation is perfect the system reaches a state of complete order where all particles share the same state, as it happens in the original MSVM.  In contrast, the addition of imperfection in the imitation rule reduces order to a level that decreases with the number of particles, leading to complete disorder in the thermodynamic limit even in the case of an infinitesimal error interval.  These conclusions are supported by two complementary analytical approaches that provide accurate expressions for the order parameters in the large population limit and in the small error amplitude limit. 

The article is organized as follows.  We introduce the model and define its dynamics in section~\ref{model}.  Section~\ref{simus} presents some simulation results showing the qualitative behavior of the model for different parameter values.  In sections~\ref{fokker-planck} and \ref{continuum} we develop two different analytical approaches that show the scaling of macroscopic quantities in different regimes.  Finally, in section~\ref{conclusions} we conclude and summarize our results.

\section{The Model}
\label{model}

\begin{figure}[t]
  \centerline{\includegraphics[width=16.0cm]{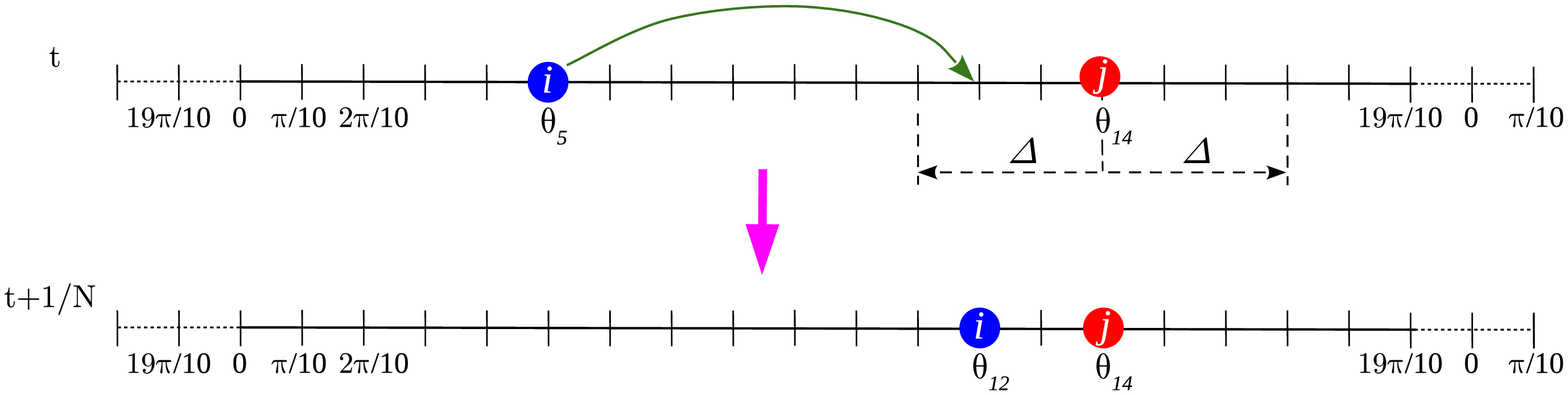}}
  \caption{Illustration of a single time step of the imperfect copying process on a chain of $S=20$ angular states labeled as $\theta_k=2 \pi k/S$ ($k=0,..,19$) with periodic boundary conditions.  Particle $i$ with state $\Theta_i(t)=\theta_5$ at time $t$ tries to imitate the state $\Theta_j(t)=\theta_{14}$ of particle $j$ by randomly jumping to one of the $7$ states contained in an interval centered at $\theta_{14}$, indicated by dashed lines.  The amplitude of the copying error interval is $\Delta=3$.  In this example, particle $i$ adopts the new state $\Theta_i (t+1/N)=\theta_{12}=\theta_{14}-2\pi/10$, similar to that of particle $j$.}  
  \label{fig-model}
\end{figure}

We consider a system of $N$ interacting particles that can take one of $S$ possible angular states $\theta_k = \frac{2 \pi k}{S}$, with $k=0,1,2,...,S-1$, which represent their moving directions.  Initially, each particle $j$ ($j=1,..,N$) adopts a state $\Theta_j=\theta_k$ at random, leading to a nearly uniform distribution of particles on a discrete angular space contained in the interval $[0,2 \pi)$.  In a single time step $\delta t=1/N$ of the dynamics, a particle $i$ is picked at random and its state $\Theta_i$ is updated according to these two steps: first, another particle $j$ with state $\Theta_j=\theta_k$ is chosen at random and, second, particle $i$ randomly adopts a state $\theta_l$ in the interval $[\theta_k - \frac{2 \pi \Delta}{S}, \theta_k + \frac{2 \pi \Delta}{S}]$ centered at $\theta_k$, i e., with equal probability $1/(1+2\Delta)$.  That is, $\Theta_i(t) \to \Theta_i(t+1/N) = \theta_l$ in the interval $[\theta_{k-\Delta},\theta_{k+\Delta}]$ (see Fig.~\ref{fig-model}).  Here $\Delta$ is a non-negative integer parameter that defines the amplitude of the error interval ($0 \le \Delta \le S/2$).  The first step corresponds to the selection of a particle $j$ whose state is tried to be imitated by particle $i$, while the second step describes the error making in the copying process, where $i$ adopts a state similar or equal to the state of $j$.  In this last step we implement periodic boundary conditions to keep the states in the $[0,2 \pi)$ interval, i e., $\theta_{-k} = \theta_{S-k}$ and $\theta_{S+k} = \theta_k$ ($0 \le k \le S-1$), and thus we can think the angle space as a chain ring of $S$ sites at positions $\theta_k$.
This dynamics can also be interpreted as a perfect copying with the addition of \emph{noise}, where particle $i$ first jumps to a site at position $\theta_k$ and then from there it jumps to any of its $2\Delta$ neighboring sites or stay in the same site with the same probability $1/(1+2\Delta)$ (see Fig.~\ref{fig-model}).

For the noiseless case $\Delta=0$ the model is equivalent to the MSVM recently studied in the literature \cite{Starnini-2012,Pickering-2016} where, in the above example, particle $i$ simply jumps to the site occupied by particle $j$ and stays there.  In this case, given that the system is only driven by the stochastic nature of the copying process (the so called genetic drift in population genetics), a site that becomes empty remains empty afterwards, as particles can jump to occupied sites only.  Therefore, the number of sites occupied by at least one particle decreases monotonically with time until only one site becomes occupied by all particles and the system stops evolving.   This configuration in which all particles share the same state --a ``consensus'' in the moving direction-- is absorbing, and thus the system has $S$ different absorbing configurations (fixation).  A magnitude of interest, which is also relevant in the analysis performed in section~\ref{continuum}, is the mean number of different states (occupies sites) in the system at time $t$, $s(t)$.  It was shown in \cite{Starnini-2012,Pickering-2016} that if $N$ particles are initially distributed homogeneously on the $S$ states [$s(0)=S$], then $s$ decays with time as 
\begin{equation}
  s(t) = \left( \frac{t}{N-1} + \frac{1}{S} \right)^{-1} ~~~
  \mbox{for $N \ge S \gg 1$}
\label{s-t}
\end{equation}
up to a time of the order $N/2$ ($s \simeq 2$), after which $s(t)$ decays exponentially fast to $1.0$.  The expected time to reach consensus can be estimated from Eq.~(\ref{s-t}) as the moment $\tau_c$ when $s(\tau_c)$ becomes $1.0$, leading to the approximate mean consensus time $\tau_c \simeq (N-1)(S-1)/S$ \cite{Starnini-2012,Pickering-2016}.

Our aim is to study how the addition of imperfection ($\Delta>0$) affects the ordering properties of the system.  We start by showing in section \ref{simus} results from Monte Carlo simulations of the dynamics, and then in sections \ref{fokker-planck} and \ref{continuum} we develop analytical approaches to gain an insight into these results.

\section{Simulation results}
\label{simus}

\begin{figure}[t]
  \vspace{0.5cm}
  \centerline{\includegraphics[width=12.0cm]{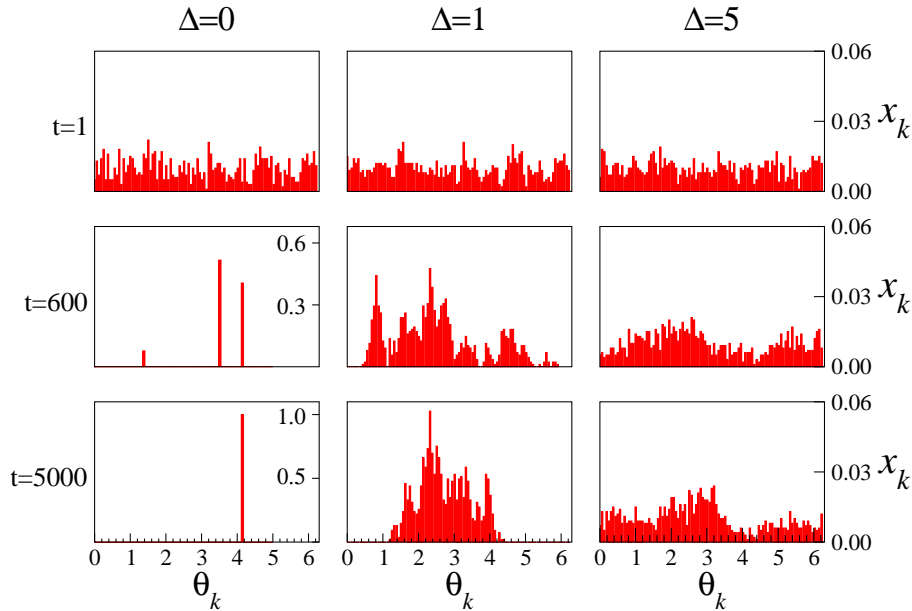}}
  \caption {Distribution of the fraction of particles $x_k$ with angular state $\theta_k$ at three different times, for $S=100$ states, $N=1000$ particles and $\Delta=0,1$ and $5$ error amplitudes.  Each three-panel column for a given value of $\Delta$ corresponds to snapshots of a single realization at times $t=1, 600$ and $5000$.}
    \label{xk-1}
\end{figure}

We simulated the dynamics of the model starting from a configuration in which each particle adopts one of the $S$ angular states $\theta_k$ at random and then evolves following the interaction rules defined in section~\ref{model}.  The state of the system at a given time $t$ can be described by the set of $S$ variables $\{x\}(t) \equiv \{x_0(t),x_1(t),..,x_{S-1}(t)\}$, where $x_k(t)$ (with $k=0,..,S-1$) is the fraction of particles with state $\theta_k=\frac{2 \pi k}{S}$ (at site $k$) at time $t$.  As the total number of particles is conserved at all times, we have $\sum_{k=0}^{S-1} x_k(t) = 1$ for all $t \ge 0$.

In order to explore how $\Delta$ affects the evolution of the system we show in Fig.~\ref{xk-1} snapshots of the distribution of the fractions $\{x\}$ at moments $t=1,600$ and $5000$ for three distinct realizations with error amplitudes $\Delta=0,1$ and $5$, for $S=100$ states and $N=10^3$ particles.  At the early time $t=1$, $\{x\}$ looks nearly uniform in all cases, but then evolves towards a distribution that depends on $\Delta$.  In the noiseless case $\Delta=0$ (left column) the system reaches a final delta distribution corresponding to a configuration where all particles are in the same state $\theta_{k=66}=4.1448$ (bottom-left panel). This is a frozen configuration where particles' states cannot longer evolve, and corresponds to one of the $S=100$ possible absorbing states of the MSVM \cite{Starnini-2012,Pickering-2016}.  Instead, for $\Delta=1$ (center column) the distribution $\{x\}$ becomes narrower with time and seems to adopt a bell shape for long times, while for $\Delta=5$ (right column) $\{x\}$ looks quite uniform for any time.  In the bottom row ($t=5000$) we observe that the width of $\{x\}$ increases with $\Delta$.  Therefore, we can see that the imperfection in the copying process is playing the role of an external noise that allows the system to escape from an absorbing configuration.

\begin{figure}[t]
  \vspace{0.5cm}
  \centerline{\includegraphics[width=12.0cm]{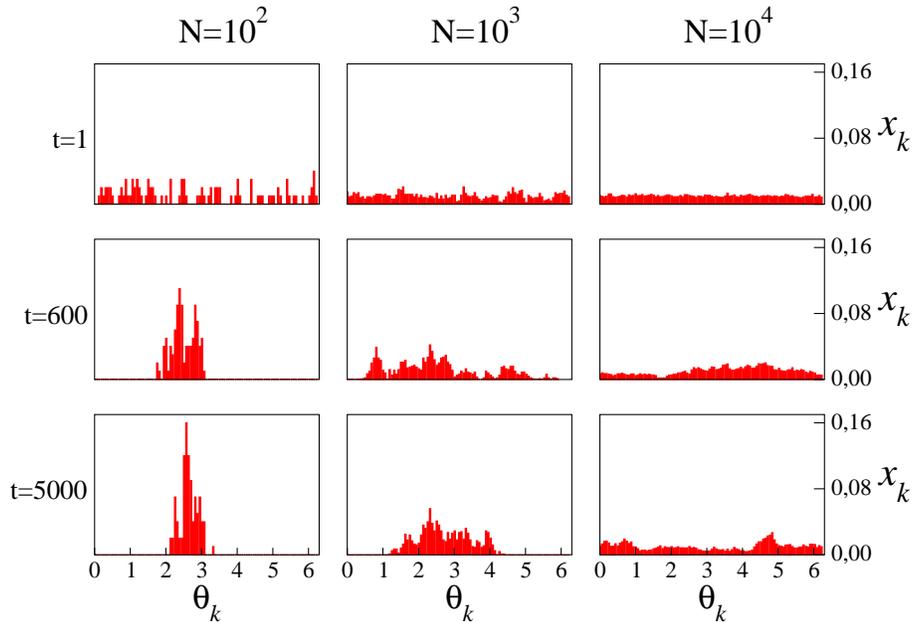}}
  \caption {Distribution of the fraction of particles $x_k$ with angular state $\theta_k$ at three different times, for $S=100$ states, error amplitude $\Delta=1$ and with $N=10^2,10^3$ and $10^4$ particles.  Each three-panel column for a given value of $N$ corresponds to snapshots of a single realization at times $t=1, 600$ and $5000$.}
  \label{xk-2}
\end{figure}

To explore the effects of varying the number of particles $N$, we show in Fig.~\ref{xk-2} the distribution $\{x\}$ at different times for $S=100$, $\Delta=1$, and system sizes $N=10^2,10^3$ and $10^4$.  We observe that for $N=10^2$ (left panels) $\{x\}$ is narrow at long times, but it becomes wider as $N$ increases (see bottom row for $t=5000$), and already looks quite uniform for large $N=10^4$.

In summary, the dynamics of the model can be roughly seen as a competition between two processes: the perfect copying of the voter dynamics that tries to bring all particles together around a single state, and the imperfect copying in the form of noise that spreads particles apart.  When $\Delta$ and $N$ are small, the system reaches a global state of order where most particles have similar angles and thus the angles' distribution is narrow, while increasing $\Delta$ and $N$ results in a wider distribution.  One may wonder how this quasi-ordered state observed for small $\Delta$ is quantitatively affected by the system size, that is, whether it reaches a stationary value as $N$ increases.  In order to investigate these issues we focus our analysis on two complementary magnitudes that characterize the system at the macroscopic level.  These are the order parameter
\begin{equation}
\psi(t) = \left| \frac{1}{N} \sum_{m=1}^N e^{i \Theta_m(t)} \right|^2 =  
  \left| \sum_{k=0}^{S-1} x_k(t) \, e^{i \theta_k} \right|^2, 
\label{psi}
\end{equation}
and the mean-squared deviation of the angular states
\begin{subequations}
  \begin{alignat}{4}
    \label{sigma}
    \sigma_{\theta}^2(t) &= \overline{\theta^2}(t) - \overline{\theta}^2(t), ~~~ \mbox{where} \\  
    \label{moments-theta}
    \overline{\theta}(t) &= \frac{1}{N} \sum_{m=1}^N \Theta_m(t)= \sum_{k=0}^{S-1} x_k(t) \, \theta_k ~~~~ \mbox{and} ~~~~
  \overline{\theta^2}(t) = \frac{1}{N} \sum_{m=1}^N \Theta^2_m(t)=\sum_{k=0}^{S-1} x_k(t) \, \theta_k^2.
\end{alignat}
\label{sigma-theta}
\end{subequations}
Here $\left| \bullet \right|$ is the absolute value, while $\Theta_m(t)$ is the state of particle $m$ ($m=1,..,N$) at time $t$.  The parameter $\psi$ ($0 \le \psi \le 1$) is similar to that introduced in the context of flocking dynamics to quantify the degree of global alignment in a system of moving particles \cite{Vicsek-1995,Baglietto-2018}, while the parameter $\sigma_{\theta}$ is a measure of the width of the distribution of angular states.  When all particles move in the same direction ($\theta_m=\theta ~ \forall m$), one can check that $\psi=1$ and $\sigma_{\theta}=0$, which corresponds to a totally ordered state.  On the other extreme, when each particle moves in a random direction the distribution of angular states becomes uniform in the $[0,2\pi)$ interval, and thus $x_k=1/S$ for $k=1,..,S-1$.  Then, defining 
$r \equiv e^{i2\pi/S}$ and writing $e^{i \theta_k}=e^{i2\pi k/S}=r^k$ the order parameter is $\psi = \left| \frac{1}{S} \sum_{k=0}^{S-1} r^k \right|^2 =\left| \frac{1-r^S}{S(1-r)} \right|^2=0$, i e., the system is completely disordered.  On its part, the mean-squared deviation takes the value 
\begin{equation}
	\sigma_{u}^2 = \frac{4 \pi^2}{S^3} \sum_{k=0}^{S-1} k^2 - \left[ \frac{2 \pi}{S^2} \sum_{k=0}^{S-1} k \right]^2 = \frac{\pi^2 (S^2-1)}{3 S^2}, 
\label{sigmau}
\end{equation}
where we have used the identities Eqs.~(\ref{sum1}) and (\ref{sum2}) in Appendix \ref{power}, with $M=S-1$, to perform the summations.

\begin{figure}[t]
  \centerline{\includegraphics[width=12.0cm]{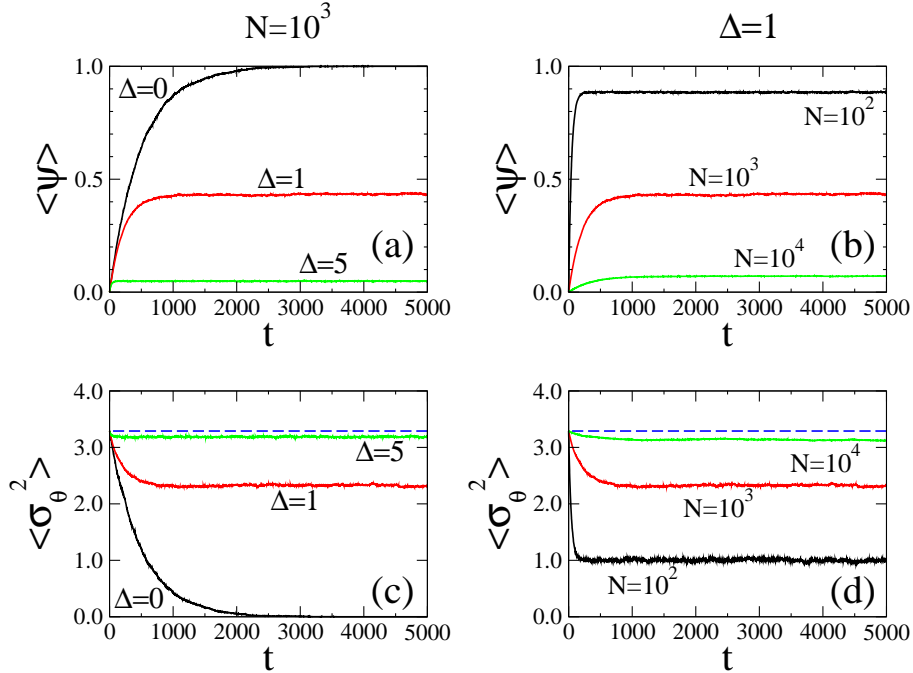}}
  \caption{Time evolution of the average value of the order parameter $\psi$ and the mean-squared deviation $\sigma_{\theta}^2$ for the same set of parameters used in Figs.~\ref{xk-1} and \ref{xk-2}, as indicated in the legends.  Panels (a) and (c) correspond to systems with $N=10^3$ particles, while panels (b) and (d) are for $\Delta=1$.  The horizontal dashed lines in panels (c) and (d) denote the mean-squared deviation $\sigma_u^2 \simeq 3.29$ of the uniform angular distribution for $S=100$ states given by Eq.~(\ref{sigmau}).  Averages were done over $10^4$ independent realizations.}
  \label{psi-sigma-t}
\end{figure}

In Fig.~\ref{psi-sigma-t} we plot the time evolution of the average value of $\psi$ and $\sigma_{\theta}^2$ over $10^4$ independent realizations, denoted by $\langle \psi \rangle$ and $\langle \sigma_{\theta}^2 \rangle$, for the same set of parameters used in Figs.~\ref{xk-1} and \ref{xk-2}.  Both magnitudes reach a stationary value that quantifies the level of order at the stationary state that corresponds to the distributions $\{x\}$ of Figs.~\ref{xk-1} and \ref{xk-2} at long times (down row).  We observe that the stationary value of $\langle \psi \rangle$ decreases monotonically with $\Delta$ [Fig.~\ref{psi-sigma-t}(a)] and $N$ [Fig.~\ref{psi-sigma-t}(b)], while the stationary value of $\langle \sigma_{\theta}^2 \rangle$ increases with $\Delta$ [Fig.~\ref{psi-sigma-t}(c)] and $N$ [Fig.~\ref{psi-sigma-t}(d)], and appears to saturate at the value of the uniform distribution $\sigma_{u}^2$ from Eq.~(\ref{sigmau}).

These results suggest that the system reaches complete order ($\psi=1$ and $\sigma_{\theta}=0$) only for the noiseless case $\Delta=0$ [Figs.~\ref{psi-sigma-t}(a) and \ref{psi-sigma-t}(c)], and that for any given error amplitude $\Delta>0$ the order constantly decreases with the system size $N$ and eventually vanishes in the thermodynamic limit [Figs.~\ref{psi-sigma-t}(b) and \ref{psi-sigma-t}(d)].  This would imply 
that a tiny amount of error in the copying dynamics is enough to lead to complete disorder ($\psi=0$ and $\sigma_{\theta}=\sigma_u$) in the $N \to \infty$ limit.  In order to analyze in more detail these conclusions obtained from numerical evidence, we develop in sections~\ref{fokker-planck} and 
\ref{continuum} two analytical approaches that allow to obtain expressions for the asymptotic behavior of $\langle \psi \rangle$ and $\langle \sigma_{\theta}^2 \rangle$ in two different limits. The approach in section~\ref{fokker-planck} is based on the diffusion approximation given by the Fokker-Planck equation and provides accurate results in the large $N$ limit, while the continuum approach developed in section~\ref{continuum} implements a superposition principle with open boundary conditions that works well in the limits of large $S$ and small noise.

\section{The Fokker-Planck approach for large $N$} 
\label{fokker-planck}

In this section we develop an analytical approach to estimate the scaling of $\psi$ and $\sigma_{\theta}^2$ with $S$, $N$ and $\Delta$ in the large $N$ limit.  We derive a Fokker-Planck equation for the distribution of particles' states that allows to obtain the behavior of the average value of 
$\psi$ and $\sigma_{\theta}^2$.  We show that, for any $\Delta>0$ and in the
$N \to \infty$ limit, the average value of the order parameter vanishes as $\langle \psi \rangle \sim 1/N$, and that the mean-squared deviation approaches the uniform value as $\sigma_u^2 - \langle \sigma_{\theta}^2 \rangle \sim 1/N$.  This result implies that even the smallest error in the copying process is enough to lead the system to complete disorder in the thermodynamic limit.  For the sake of simplicity, we focus on the simplest non-trivial case $\Delta=1$ 
where, in an iteration step, a randomly chosen particle tries to copy the state $\theta_j$ of another random particle, adopting either state $\theta_{j-1}$, $\theta_j$ or $\theta_{j+1}$ with equal probability $1/3$.  We then use some heuristic arguments to extend these results to the general case $\Delta >1$.

Even though we are aware that there might be many different ways to address this problem analytically, we follow here a physics approach based on the diffusion approximation that gives the Fokker-Planck equation.  This approach is particularly useful in the $N >> 1$ limit because it allows to obtain rather accurate expressions for the stationary second moments $\langle x_i x_j \rangle$ that appear in the average values of both $\psi$ and $\sigma_{\theta}^2$ when we expand Eqs.~(\ref{psi}) and (\ref{sigma}), respectively.

We start by describing the state of the system by the set of variables \\
$\{x\}=\{x_0,x_1,..,x_{S-1}\}$, where $x_k$ ($k=0,..,S-1$) is the fraction of particles with state $\theta_k$ (at site $k$) subject to the constraint $\sum_{k=0}^{S-1} x_k=1$ for all times.  When a particle makes a transition from site $k$ to site $j \ne k$, the state of the system changes from $\{x\}$ to a new state denoted by
$ \{x'\}_{k \; j}^{- +} \equiv \{x_0,..,x_k-1/N,..,x_j+1/N,..,x_{S-1}\}$ in which, compared to $\{x\}$, site $k$ has lost a particle ($x_k \to x_k-1/N$) and site $j$ has gained a particle
($x_j \to x_j+1/N$). This is indicated in the notation $\{x'\}_{k \; j}^{- +}$ with the $-$ and $+$ signs on top of subindices $k$ and $j$, respectively.
The probability $P(\{x\},t)$ that the system is in state $\{x\}$ at time $t$ obeys the master equation
\begin{equation}
  \frac{d}{d t} P(\{x\},t)=\sum_{k=0}^{S-1} \sum_{\substack{j=0 \atop j \ne k}}^{S-1}
  \Big\{ W \left( \{x'\}_{k \; j}^{- +} \to \{x\} \right) \, P \left( \{x'\}_{k \; j}^{- +},t \right) -  W \left( \{x\} \to \{x'\}_{k \; j}^{- +} \right) \, P \left( \{x\},t \right) \Big\}. 
\label{me}
\end{equation}
The transition rate $W \left( \{x\} \to \{x'\}_{k \; j}^{- +} \right)$ is the
probability per time step $\delta t=1/N$ that a particle jumps from site $k$ to site $j$, calculated as $\frac{N}{3} x_k (x_{j-1}+x_j+x_{j+1})$.  That is, a particle in site $k$ is chosen with probability $x_k$, then it jumps to either sites $j-1$, $j$ or $j+1$ with probability $x_{j-1}+x_j+x_{j+1}$, and from there jumps to site $j$ with probability $1/3$.  According to the periodic character of angles, we make $x_{-1}=x_{S-1}$ and $x_S=x_0$ for the $j=0$ and $j=S-1$ cases, respectively.  Similarly, we can calculate the transition rate $W \left( \{x'\}_{k \; j}^{- +} \right) \to \{x\} )$ that corresponds to a particle that jumps from site $j$ to site $k$.  Then, the transition rates are given by the expressions:
\begin{eqnarray}
 W_{k \, j}^{\downarrow \uparrow} \left(\{x\} \right) &\equiv& W \left( \{x\} \to \{x'\}_{k \; j}^{- +} \right) = \frac{N}{3} x_k (x_{j-1}+x_j+x_{j+1}) ~~~ \mbox{and} \nonumber \\    
 W_{k \, j}^{\uparrow \downarrow} \left(\{x\} \right) &\equiv&
W \left( \{x\} \to \{x'\}_{k \; j}^{+ -} \right) = \frac{N}{3} x_j (x_{k-1}+x_k+x_{k+1}), ~~~ \mbox{for $S \ge 3$,}
\label{rates}
\end{eqnarray}
and
\begin{eqnarray}
  W_{0 \, 1}^{\downarrow \uparrow} \left(\{ x_0,x_1 \} \right) &\equiv& W \left( \{x_0,x_1\} \to \{x_0-1/N,x_1+1/N\} \right) = \frac{N}{2} x_0  ~~~ \mbox{and} \nonumber \\    
 W_{0 \, 1}^{\uparrow \downarrow} \left(\{x_0,x_1\} \right) &\equiv&
W \left(\{x_0,x_1\} \to \{x_0+1/N,x_1-1/N\} \right) = \frac{N}{2} x_1, ~~~ \mbox{for $S=2$}.
\label{rates-2}
\end{eqnarray}
For convenience, we have simplified notation using the rising and lowering operators $W_{k \, j}^{\downarrow \uparrow}$ and $W_{k \, j}^{\uparrow \downarrow}$.  For instance, the down (up) arrow on top of $k$ ($j$) indicates that the operator applied on $\{x\}$ decreases (increases) $x_k$ ($x_j$) in $1/N$.  The transitions for the case $S=2$ were displayed separately because they take a different form.  In this particular case there are only two angular states, $\theta_0=0$ and $\theta_1=\pi$, and thus the noise step moves a particle to any of the two angles with equal probability $1/2$, instead of probability $1/3$ as explained above for any $S \ge 3$. 

We can now obtain the Fokker-Planck equation by Taylor expanding the first term of Eq.~(\ref{me}) up to second order in $1/N \ll 1$ for large $N$:
\begin{eqnarray}
 W_{k \, j}^{\uparrow \downarrow} \left(\{x'\}_{k \; j}^{- +} \right) P \left( \{x'\}_{k \; j}^{- +},t \right) &=& W_{k \, j}^{\uparrow \downarrow} \, P + \frac{1}{N} \left( \frac{\partial}{\partial x_j} - \frac{\partial}{\partial x_k} \right) \left[ W_{k \, j}^{\uparrow \downarrow} \, P \right] \nonumber \\ &+& \frac{1}{2N^2} \left( \frac{\partial^2}{\partial x_j^2} + \frac{\partial^2}{\partial x_k^2} - \frac{2 \, \partial^2}{\partial x_k \partial x_j} \right) \left[ W_{k \, j}^{\uparrow \downarrow} \, P \right] + \mathcal O(1/N^3),
\label{taylor}
\end{eqnarray}
where $W_{k \, j}^{\uparrow \downarrow}$ and $P$ are short notations for
$W_{k \, j}^{\uparrow \downarrow}(\{x\})$ and $P(\{x\},t)$, respectively, which are the functions $W$ and $P$ applied to the unperturbed state $\{x\}$.  Inserting expression Eq.~(\ref{taylor}) into Eq.~(\ref{me}) leads to 
\begin{eqnarray}
  \partial_t P( \{x\},t) &=& - \frac{1}{N} \sum_{k=0}^{S-2} \partial_k \Bigg\{ \sum_{\substack{j=0 \atop j \ne k}}^{S-1}  \left[ W_{k \, j}^{\uparrow \downarrow}(\{x\}) -
    W_{k \, j}^{\downarrow \uparrow}(\{x\}) \right] P(\{x\},t) \Bigg\} \nonumber \\
  &+& \frac{1}{2N^2} \sum_{k=0}^{S-2} \partial_{k k}^2 \Bigg\{
\sum_{\substack{j=0 \atop j \ne k}}^{S-1} \left[ W_{k \, j}^{\uparrow \downarrow}(\{x\}) +
  W_{k \, j}^{\downarrow \uparrow}(\{x\}) \right] P(\{x\},t) \Bigg\} \nonumber \\ 
&-& \frac{1}{N^2} \sum_{k=0}^{S-3} \sum_{j > k}^{S-2} \partial_{k j}^2 \Bigg\{
\left[ W_{k \, j}^{\uparrow \downarrow}(\{x\}) +
  W_{k \, j}^{\downarrow \uparrow}(\{x\}) \right] P(\{x\},t) \Bigg\},
\label{fpe-1}
\end{eqnarray}
where $\partial_k \equiv \partial/\partial x_k$ and $\partial_{k j}^2 \equiv \partial^2/\partial x_k \partial x_j$.  To arrive to Eq.~(\ref{fpe-1}) we have made two considerations.  First, we have used the following equalities to simplify the summations:
\begin{eqnarray}
  \sum_{k=0}^{S-1} \sum_{\substack{j=0 \atop j \ne k}}^{S-1} \partial_j
  \left( W_{k \, j}^{\uparrow \downarrow} \, P \right) &=&
  \sum_{k=0}^{S-1} \partial_k \Bigg\{ \sum_{\substack{j=0 \atop j \ne k}}^{S-1} 
  W_{k \, j}^{\downarrow \uparrow} \, P \Bigg\} ~~~\mbox{and} \nonumber \\
  \sum_{k=0}^{S-1} \sum_{\substack{j=0 \atop j \ne k}}^{S-1} \partial_{k j}^2
  \left( W_{k \, j}^{\uparrow \downarrow} \, P \right) &=&
  \sum_{k=0}^{S-2} \sum_{j > k}^{S-1} \partial_{k j}^2 \Big\{
  \left( W_{k \, j}^{\uparrow \downarrow} + W_{k \, j}^{\downarrow \uparrow} \right) P \Big\}. \nonumber   
\end{eqnarray}
Second, we have used the constraint $\sum_{k=0}^{S-1} x_k =1$ to write $x_{S-1}$ in terms of the other fractions, $x_{S-1}=1-\sum_{k=0}^{S-2} x_k$, reducing the number of independent variables to $S-1$.  This makes partial derivatives $\partial/\partial x_{S-1}$ vanish, and set to $S-2$ the upper limit of the summation over $k$.

Plugging expressions from Eq.~(\ref{rates}) and Eq.~(\ref{rates-2}) for the transition rates into Eq.~(\ref{fpe-1}), and performing the summations inside the brackets we arrive to the Fokker-Planck equation in its final form
\begin{eqnarray}
\partial_t P( \{x\},t) &=& - \sum_{k=0}^{S-2} \partial_k \left[ A_k \,
    P( \{x\},t) \right] + \frac{1}{2} \sum_{k=0}^{S-2} \partial_{k k}^2 \left[ B_{kk} \, P( \{x\},t) \right] \nonumber \\ &+& \sum_{k=0}^{S-3} \sum_{j > k}^{S-2} \partial_{k j}^2 \left[ B_{kj} \, P(\{x\},t) \right], ~~ \mbox{for $S \ge 3$},
\label{fpe-2}
\end{eqnarray}
where 
\begin{eqnarray}
\label{ABk}
  A_k &=& \frac{1}{3} ( x_{k-1} -2 x_k + x_{k+1} ), \nonumber \\
  B_{kk} &=& \frac{1}{3N} \left[ 2 x_k (2-x_k) + (1-2 x_k)(x_{k-1} + x_{k+1}) \right], \\  
  B_{kj} &=& - \frac{1}{3N} \left[ x_k (x_{j-1}+x_{j+1}) + x_j (x_{k-1} + x_{k+1}) + 2 x_k x_j \right]. \nonumber
\end{eqnarray}
and 
\begin{equation}
  \frac{\partial}{\partial t} P(x_0,t) = \frac{1}{2} \frac{\partial}{\partial x_0} \left[(2x_0-1) P(x_0,t) \right] + \frac{1}{4N} \frac{\partial^2}{\partial x_0^2} P(x_0,t), ~~ \mbox{for $S=2$}.
\label{fpe-3}
\end{equation}
Equations~(\ref{fpe-2}), (\ref{ABk}) and (\ref{fpe-3}) give the time evolution of the probability distribution of angular states in a population of $N$ particles.  The stationary solution of this Fokker-Planck equation, denoted by $P_{st}(\{x\})$, can be used to obtain the average value of $\psi$ and $\sigma_{\theta}^2$ at the stationary state, as we do in the following subsections.

\subsection{Analysis of the $S=2$ case}
\label{S2}

In order to gain an analytical insight into the behavior of the system at the stationary state we start by studying the simplest case of two angular states $S=2$ ($\theta=0,\pi$).  We notice that this $2$-state model corresponds to a particular case of a surface-reaction model with noise studied in \cite{Considine-1989} where, in a single step of the dynamics, one randomly chosen particle takes either state $1$ of $-1$ with probability $p_d/2$, or copies the state of a random neighbor with the complementary probability $1-p_d$.  When $p_d=1$ the surface-reaction model turns equivalent to our model for $S=2$.  Equation~(\ref{fpe-3}) describes the time evolution of the probability of finding a fraction $x_0$ of particles with angle $\theta_0=0$, whose stationary solution with boundary conditions $P_{st}(0)=P_{st}(1)$ and
$\left. \frac{\partial P_{st}}{\partial x_0} \right|_{x_0=0}=\left. -\frac{\partial P_{st}}{\partial x_0} \right|_{x_0=1}$ is 
\begin{equation}
  P_{st}(x_0) = \frac{\sqrt{2 N} \; e^{-2 N (x_0-1/2)^2}}{\sqrt{\pi} \;  \mbox{erf} \left(\sqrt{\frac{N}{2}}\right)},
\label{Pst}
\end{equation}
which satisfies the normalization condition $\int_0^1 P_{st}(x_0) \, dx_0=1$.  One can check that expression Eq.~(\ref{Pst}) corresponds to the $p_d \to 1$ limit of the stationary solution found in \cite{Considine-1989}.  The reason why we assumed these particular boundary conditions for $P_{st}$ is because both states $\theta_0=0$ and $\theta_1=\pi$ are equivalent, and thus we expect $P_{st}$ to be symmetric around $x_0=1/2$.  We see that the stationary distribution of the fraction of particles with angle $\theta_0=0$ given by Eq.~(\ref{Pst}) is a Gaussian centered at $x_0=1/2$, whose width decreases as $N^{-1/2}$ with the number of particles.  The order parameter from Eq.~(\ref{psi}) becomes
\begin{equation}
  \psi(t) = \left| x_0(t)-x_1(t) \right|^2 = \left[ 2x_0(t)-1 \right]^2.
  \label{psi-t}
\end{equation}
Then, the average value of $\psi$ at the stationary state can be calculated using $P_{st}(x_0)$ from Eq.~(\ref{Pst}) as
\begin{eqnarray}
\langle \psi \rangle &=& \int_0^1 (2x_0-1)^2 \, P_{st}(x_0) \, dx_0 \nonumber \\ &=& \frac{\sqrt{2 N}}{\sqrt{\pi} \; \mbox{erf} \left(\sqrt{\frac{N}{2}}\right)}
    \, \int_{-1}^1 y^2 \, e^{-\frac{N}{2} y^2} dy = 
  \frac{1}{N} - \frac{\sqrt{2} \, e^{-N/2} }{\sqrt{\pi N} \, \mbox{erf} \left(\sqrt{\frac{N}{2}}\right)}, 
  \label{psiave-1}
\end{eqnarray}
where we have made the change of variables $y=2x_0-1$ and integrated by parts.  To first order in $1/N$, Eq.~(\ref{psiave-1}) is reduced to the simple expression
\begin{eqnarray}
  \langle \psi \rangle \simeq \frac{1}{N},
  \label{psiave-2}
\end{eqnarray}
which shows that $\langle \psi \rangle$ vanishes in the $N \to \infty$ limit.  On its part, the mean-squared deviation from Eqs.~(\ref{sigma-theta}) is 
\begin{equation}
  \sigma_{\theta}^2(t)=\pi^2 x_1(t) [1-x_1(t)] = \pi^2 x_0(t) [1-x_0(t)],
  \label{sigma-t}
\end{equation}
and thus its stationary average value is calculated as 
\begin{eqnarray}
 \langle \sigma_{\theta}^2 \rangle &=& \pi^2 \int_0^1 x_0 (1-x_0) \, P_{st}(x_0) \, dx_0 \nonumber \\  &=& \frac{\pi^2 \sqrt{2 N}}{8 \sqrt{\pi} \; \mbox{erf} \left(\sqrt{\frac{N}{2}}\right)}
  \, \int_{-1}^1 (1-y^2) \, e^{-\frac{N}{2} y^2} dy = \frac{\pi^2}{4}
  \left[ 1 - \frac{1}{N} + \frac{\sqrt{2} \, e^{-N/2} }{\sqrt{\pi N} \, \mbox{erf} \left(\sqrt{\frac{N}{2}}\right)} \right].
\label{sigma-ave-1}
\end{eqnarray}
For $N \gg 1$, Eq.~(\ref{sigma-ave-1}) is reduced to the simple expression
\begin{equation}
  \langle \sigma_{\theta}^2 \rangle \simeq \frac{\pi^2}{4} \left(1-\frac{1}{N}  \right) = \sigma_u^2 \left(1-\frac{1}{N} \right),
\label{sigma-ave-2}
\end{equation}
which shows that $\langle \sigma_{\theta}^2 \rangle$ approaches the mean-squared deviation $\sigma_u^2=\pi^2/4$ of the uniform distribution as $N \to \infty$.  Equations~(\ref{psiave-2}) and (\ref{sigma-ave-2}) describe the main result in the analysis of ordering in the noisy MSVM, that is, the distribution of angular states becomes uniform in the $N \to \infty$, and thus the system achieves total disorder ($\psi=0$).  Even though this applies here only for the two--angle case $S=2$, we shall see in the next subsection that the same scalings with $N$ are obtained for any $S \ge 3$ as well.

To interpret these results from the dynamics of the system we resort to Eq.~(\ref{Pst}) and observe that, at the stationary state of a single realization, $x_0$ and $x_1$ fluctuate around the value $1/2$ subject to the constraint $x_0(t)+x_1(t)=1$.  When $N$ increases, the amplitude of fluctuations vanishes as $N^{-1/2}$, and thus $P_{st}$ tends to the delta function $P_{st}(x_0)=\delta(x_0-1/2)$.  Therefore, we obtain the expected results $\psi=0$ and $\sigma_{\theta}^2 = \sigma_u^2 = \pi^2/4$ from Eqs.~(\ref{psi-t}) and (\ref{sigma-t}), respectively.

\subsection{Analysis of the general case $S \ge 2$}
\label{S3}

In the last section we obtained expressions for $\langle \psi \rangle$ and $\langle \sigma_{\theta}^2 \rangle$ when $S=2$ from the stationary solution Eq.~(\ref{Pst}) of the Fokker-Planck equation.  However, it seems hard to integrate analytically Eq.~(\ref{fpe-2}) and find an expression for the stationary distribution $P_{st}(\{x\})$ for the general case $S \ge 3$.  Nevertheless, we shall see in the next two subsections that $\langle \psi \rangle$ and 
$\langle \sigma_{\theta}^2 \rangle$ can be estimated by expressing them in terms of the second moments of $P_{st}(\{x\})$ which, in the limit of large $N$, can be obtained without knowing the explicit functional form of $P_{st}(\{x\})$.

\subsubsection{Calculation of the order parameter $\langle \psi \rangle$}
\label{section-psi}

We start by using the equality $e^{i \theta_k} = \cos \theta_k + i \sin \theta_k$ and rewriting the order parameter from Eq.~(\ref{psi}) as
\begin{equation}
  \psi(t) = \left( \sum_{k=0}^{S-1} x_k(t) \cos \theta_k \right)^2 +  \left( \sum_{k=0}^{S-1} x_k(t) \sin \theta_k \right)^2.
\label{phi}
\end{equation}
Expanding the two squared terms of Eq.~(\ref{phi}) leads to  
\begin{eqnarray*}
 \psi(t) &=& \sum_{k=0}^{S-1} x_k^2(t) \left( \cos^2 \theta_k + \sin^2 \theta_k \right) + 2 \sum_{k=0}^{S-2} \sum_{j>k}^{S-1} x_k(t) \, x_j(t) \left( \cos \theta_k \cos \theta_j + \sin \theta_k \sin \theta_j \right) \nonumber \\ 
  &=& \sum_{k=0}^{S-1} x_k^2(t) + 2 \sum_{k=0}^{S-2} \sum_{j>k}^{S-1} x_k(t) \, x_j(t) \cos\left[2 \pi (j-k)/S \right],   
\end{eqnarray*}
where we have used the formula for the cosine of the sum of two angles.  Now, the average value of $\psi$ at the stationary state is 
\begin{eqnarray}
\langle \psi \rangle &=& \sum_{k=0}^{S-1} \langle x_k^2 \rangle + 2 \sum_{k=0}^{S-2} \sum_{j>k}^{S-1} \langle x_k x_j \rangle \cos\left[2 \pi (j-k)/S \right] \nonumber \\ &=&  S z_0 + 2 \sum_{k=0}^{S-2} \sum_{j>k}^{S-1} z_{j-k} \cos\left[2 \pi (j-k)/S \right]. 
\label{psi-1}
\end{eqnarray}
Here we have exploited the translational symmetry of the angle's space (a chain ring).  We assumed that the second moments are invariant under translation, and thus they are a function $z_{j-k}$ of the distance $j-k \ge 0$ between $j$ and $k$, i e., $\langle x_k x_j \rangle = \langle x_{k-1} x_{j-1} \rangle = \langle x_{k-2} x_{j-2} \rangle =...=\langle x_0 x_{j-k} \rangle \equiv z_{j-k}$.  In particular, we have $\langle x_k^2 \rangle = \langle x_0^2 \rangle \equiv z_0$ for all $k=0,..S-1$.  Then, expressing the double sum of the second term of Eq.~(\ref{psi-1}) as a single sum over the index $n \equiv j-k \ge 0$, we arrive to
\begin{eqnarray}
  \langle \psi \rangle = S z_0 + 2 \sum_{n=1}^{S-1} (S-n) z_n \cos(2 \pi n/S).    
\label{psi-2}
\end{eqnarray}
To perform the summation in Eq.~(\ref{psi-2}) we need to calculate the stationary value of the moments $z_n = \langle x_k x_j \rangle$.  Given that we do not know how to obtain $P_{st}$ for 
$S \ge 3$ we use a different approach, developed in Appendix \ref{moments}.  That is, starting from the Fokker-Planck equation we derive a system of coupled difference equations that relate the moments $z_n$ at the stationary state, whose solution gives the following approximate expressions for $z_n$ in the $N \gg S^2 > 1$ limit (see Appendix \ref{moments} for calculation details):
\begin{subequations}
  \begin{alignat}{4}
    \label{zn2}
    z_0 & \simeq \frac{1}{4} \left(1+\frac{1}{N} \right) ~~~ \mbox{and} ~~~ 
    z_1 \simeq \frac{1}{4} \left(1-\frac{1}{N} \right) ~~~ \mbox{for $S=2$}, ~~~ \mbox{and} \\
    \label{zn3}
    z_n &\simeq \frac{1}{S^2} \left[1-\frac{1-S^2 + 6 n (S-n)}{4N} \right], ~ \mbox{with $n=0,..,S-1$, for $S \ge 3$}.
  \end{alignat}
  \label{zn}
\end{subequations}
Inserting expressions Eqs.~(\ref{zn2}) for $z_0$ and $z_1$ into Eq.~(\ref{psi-2}) we obtain for $S=2$ 
\begin{eqnarray}
  \langle \psi \rangle = 2 (z_0 -z_1) \simeq \frac{1}{N} \nonumber,  
\end{eqnarray}
which agrees with the expression Eq.~(\ref{psiave-2}) obtained in section~\ref{S2} by direct calculation in the $N \gg 1$ limit.  Now, for $S \ge 3$ we plug expression Eq.~(\ref{zn3}) for $z_n$ into Eq.~(\ref{psi-2}) and find, after doing some algebra,
\begin{eqnarray}
  \label{psi-3}
  \langle \psi \rangle &\simeq& \frac{1}{S^2} \left( S+2 \, a_S \right) \left(1-\frac{1-S^2}{4N} \right) - \frac{3 \, b_S}{S^2 N}, ~~~ \mbox{with} \\
  \label{as0}
  a_S &=& \sum_{n=1}^{S-1} (S-n) \cos(2 \pi n/S) ~~~ \mbox{and} \\
  \label{bs0}
  b_S &=& \sum_{n=1}^{S-1} n (S-n)^2 \cos(2 \pi n/S).
\end{eqnarray}
The summations above can be calculated exactly using complex variables (see Appendix \ref{coefficients} for details), obtaining
\begin{eqnarray}
  \label{as}
  a_S &=& -\frac{S}{2} ~~~ \mbox{and} \\
  \label{bs}
  b_S &=& - \frac{S^2}{4 \sin^2(\pi/S)}.  
\end{eqnarray}
Finally, inserting these expressions for the coefficients $a_S$ and $b_S$ into Eq.~(\ref{psi-3}) we arrive to the following approximate expression for the order parameter
\begin{eqnarray}
  \langle \psi \rangle \simeq \frac{3}{4 \sin^2(\pi/S) \, N} ~~~~~ \mbox{for $\Delta=1$, $S \ge 3$ and $N \gg S^2$.}
\label{psi-4}
\end{eqnarray}
This result shows that for any $S \ge 3$ the order parameter vanishes as $1/N$ in the $N \to \infty$ limit.  

\begin{figure}[t]
  \vspace{0.5cm}
  \centerline{\includegraphics[width=8.0cm]{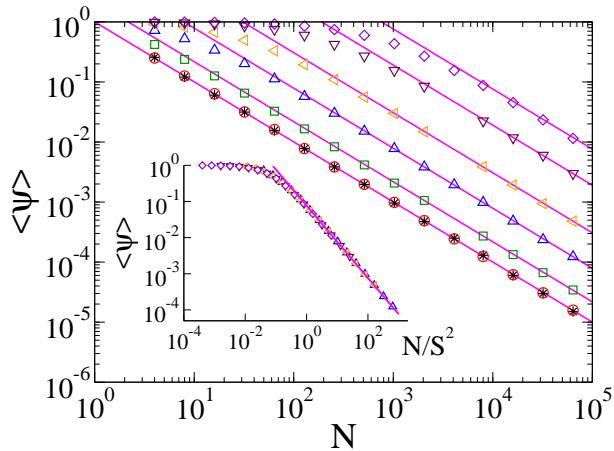}}
  \caption{Average value of the order parameter $\psi$ at the stationary state vs system size $N$ for error amplitude $\Delta=1$ and number of angular states $S=2$ (stars), $S=3$ (circles), $S=5$ (squares), $S=10$ (up triangles), $S=20$ (left triangles), $S=50$ (down triangles) and $S=100$ (diamonds).  Straight lines are the theoretical approximation given by Eq.~(\ref{psi-4}).  The inset shows the collapse of the data when the $x$-axis is rescaled by $S^2$, and that the analytical approximation from Eq.~(\ref{psi-5}) (straight line) is valid for $N \gtrsim S^2$.  Averages were done over $10^4$ independent realizations.}
  \label{psi-N-delta1}
\end{figure}
 
In Fig.~\ref{psi-N-delta1} we compare the behavior of $\langle \psi \rangle$ from Eq.~(\ref{psi-4}) (solid lines) with that obtained from Monte Carlo (MC) simulations for $\Delta=1$ and different values of $S$ and $N$ (symbols).  We observe that, for a given $S$, the agreement between the analytical curve and
the numerical data becomes better as $N$ increases, and it is very good when
$N \gtrsim S^2$.  This suggests plotting the data as a function of the rescaled variable $u \equiv N/S^2$, as we show in the inset of Fig.~\ref{psi-N-delta1}.  The straight solid line is the analytical approximation 
\begin{equation}
  \langle \psi \rangle(u) \simeq \frac{3}{4\pi^2 \, u} ~~~ \mbox{for  $S \gg 1$},
\label{psi-5}
\end{equation}
obtained by expanding Eq.~(\ref{psi-4}) to first order in $1/S$.  We see that all data collapses into a single curve that follows the power law decay Eq.~(\ref{psi-5}) when $u$ is approximately larger than $1$, i e., for $N \gtrsim S^2$ as mentioned above.

Even though the above analysis was performed for the particular case in which the angle perturbation is to first nearest-neighbor only ($\Delta=1$), we shall see that similar scalings hold for $\Delta > 1$.  In the general case $\Delta \ge 1$, it proves useful to consider each iteration of the dynamics as the two-step process (copy $+$ noise) described in section~\ref{model}, where the noise is represented by a uniform random variable $\xi$ that takes a value in the discrete set $-2 \pi \Delta/S, -2 \pi (\Delta-1)/S, .., 0, .., 2 \pi \Delta/S$ with the same probability $1/(1+2\Delta)$.  In order to generalize expression Eq.~(\ref{psi-4}) for $\Delta>1$ we shall assume that $\langle \psi \rangle$ is a function of the noise variance $\sigma_{\xi}$, calculated as
\begin{eqnarray*}
  \sigma_{\xi}(\Delta,S)=\sqrt{ \langle \xi^2 \rangle - \langle \xi \rangle^2}= \frac{\pi}{\sqrt 3} \, \eta(\Delta,S),
\end{eqnarray*}
where we have defined
\begin{equation}
  \eta(\Delta,S) \equiv \frac{2\Delta}{S} \sqrt{1+1/\Delta}.
  \label{eta-Delta-S}
\end{equation}
Note that by letting $\Delta$ and $S$ go to infinity while keeping the ratio $\Delta/S$ fixed, $\eta$ reduces to the simple expression $\eta=2 \Delta/S$ that is the noise amplitude in the case of continuum angles $\theta \in [0,2\pi)$ when $S \to \infty$ (we shall exploit this observation in section~\ref{continuum}).  For $\Delta=1$, we can write $S$ in terms of $\eta$ from Eq.~(\ref{eta-Delta-S}) as $S=2 \sqrt{2}/\eta$.  Then, replacing this expression for $S$ into Eq.~(\ref{psi-4}) we obtain
\begin{equation}
  \langle \psi \rangle \simeq \frac{3}{4 \sin^2 \left( \frac{\pi \, \eta}{2 \sqrt{2}} \right) N}, ~~~ \mbox{for $S \ge 3$ and $N \gg 1/\eta^2$.}
\label{psi-eta}
\end{equation}
with $\eta=\eta(\Delta,S)$ given by Eq.~(\ref{eta-Delta-S}).

\begin{figure}[t]
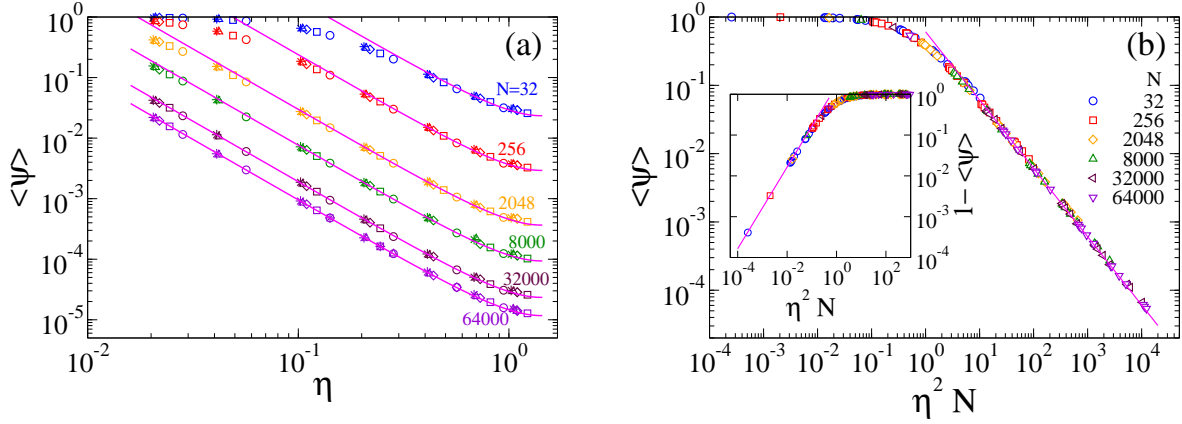

  \begin{center}
    \vspace{0.5cm}
    \begin{tabular}{cc}
      \hspace{-1.0cm}
      \includegraphics[width=5.0cm, bb=70 -20 550 550]{Fig6a.eps}
      & \hspace{3.0cm} 
      \includegraphics[width=5.0cm, bb=70 -20 550 550]{Fig6b.eps} 
    \end{tabular}    
    \caption{(a) Average value of $\psi$ vs relative error amplitude $\eta(\Delta,S)$ for system sizes $N=32$ (blue symbols), $N=256$ (red symbols), $N=2048$ (orange symbols), $N=8000$ (green symbols), $N=32000$ (maroon symbols) and $N=64000$ (violet symbols).  The values of the error amplitude are $\Delta=1$ (circles), $\Delta=2$ (squares), $\Delta=5$ (diamonds), $\Delta=10$ (triangles) and $\Delta=20$ (stars), while the values of $S$ used for each $\Delta$ were chosen to give $\eta$ in the range $[0.02,1.23]$ according to Eq.~(\ref{eta-Delta-S}).  Continuous curves correspond to the analytical estimation Eq.~(\ref{psi-eta}) for general $N$, $\Delta$ and $S$.  (b) Collapse of the data points shown in panel (a) when they are plotted as a function of the rescaled variable $\eta^2 N$.  The straight line is the analytical approximation from Eq.~(\ref{psi-6}) which shows that $\langle \psi \rangle$ decays as $6/(\pi^2 \eta^2 N)$ for $\eta^2 N \gtrsim 1$.  Inset:  $\langle \psi \rangle$ approaches $1$ when $\eta^2 N$ approaches zero.  The straight line is the approximation $c_s \, \eta^2 N$ from Eq.~(\ref{psi-low-3}), with $c_s=1.64$ (see section~\ref{section-psi-2}).}
    \label{psi-eta-all}
  \end{center}
\end{figure}

To test the validity of the relation between $\langle \psi \rangle$ and $\eta$ from Eq.~(\ref{psi-eta}) for any $\Delta$ and $S$ we have performed simulations for $\Delta=1,2,5,10,20$, seven values of $S$ for each $\Delta$, and various system sizes.  Results are shown in Fig.~\ref{psi-eta-all}(a) (symbols) where we plot $\langle \psi \rangle$ vs $\eta$, with $\eta$ given by Eq.~(\ref{eta-Delta-S}) for each pair $(\Delta,S)$. Note that, for fixed values of $N$ and $\Delta$, changing $S$ implies varying $\eta$ along the $x$-axis.  Each of the solid curves corresponds to the analytical prediction Eq.~(\ref{psi-eta}) for a fixed value of $N$ and varying $\eta$ continuously in the range $[0.01,1.44]$.  We observe that expression Eq.~(\ref{psi-eta}) is a good estimation of the numerical value of $\langle \psi \rangle$ within a range of $\eta$ that increases with $N$.  This shows that for any $\Delta \ge 1$, $S \ge 2 \Delta$ and $N$ large, $\langle \psi \rangle$ can be expressed as a function of the parameter $\eta(\Delta,S)$ given by Eq.~(\ref{eta-Delta-S}).  Analogously to the $\Delta=1$ case, we can expand Eq.~(\ref{psi-eta}) to first order in $\eta$ and obtain 
\begin{equation}
  \langle \psi \rangle \simeq \frac{6}{\pi^2 \eta^2 N} ~~~ \mbox{for $\eta \ll 1$ and $N \gtrsim 1/\eta^2$},
\label{psi-6}
\end{equation}
which suggests the scaling $\langle \psi \rangle=f(\eta^2 N)$.  Indeed, we can see in Fig.~\ref{psi-eta-all}(b) a good data collapse when the data is plotted as a function of $\eta^2 N$ (for $\eta < 0.6$), and that obeys the power law decay from Eq.~(\ref{psi-6}) (solid line) when $\eta^2 N \gtrsim 1$.

\subsubsection{Calculation of the mean-squared deviation $\langle \sigma_{\theta}^2 \rangle$}
\label{section-sigma}

From the definition Eqs.~(\ref{sigma-theta}), the mean-squared deviation can be expressed as 
\begin{eqnarray*}
 \sigma_{\theta}^2(t) = \sum_{k=0}^{S-1} x_k(t) \, \theta_k^2 - \left[ \sum_{k=0}^{S-1} x_k(t) \, \theta_k \right]^2  = \frac{4 \pi^2}{S^2} \sum_{k=0}^{S-1} k^2 x_k(t) \left[ 1-x_k(t) \right] - \frac{8 \pi^2}{S^2} \sum_{k=0}^{S-2} \sum_{j>k}^{S-1} k \, j \, x_k(t) \, x_j(t). 
\end{eqnarray*}
Then, the average value $\sigma_{\theta}^2$ at the stationary state is 
\begin{eqnarray}
    \langle \sigma_{\theta}^2 \rangle = \frac{2 \pi^2}{3 S^2} \left( 1 - S z_0 \right) (S-1) (2S-1) - 
    \frac{8 \pi^2}{S^2} \sum_{k=0}^{S-2} \sum_{j>k}^{S-1} k \, j \, z_{j-k}, 
    \label{sigma-ave-3}
\end{eqnarray}
where we have replaced $\langle x_k \rangle$ by $1/S$ [see Eq.~(\ref{xk1S}) in Appendix \ref{moments} for calculation details], $\langle x_k x_j \rangle$ by $z_{j-k}$ and expressed the summation $\sum_{k=0}^{S-1} k^2$ in terms of $S$ using Eq.~(\ref{sum2}).  The double summation in Eq.~(\ref{sigma-ave-3}) can we rewritten in terms of the index $n=j-k$ as 
\begin{eqnarray*}
\sum_{n=1}^{S-2} z_n \sum_{k=1}^{S-1-n} k(k-n) = \frac{1}{6} \sum_{n=1}^{S-2} (S-1-n)(S-n)(2S-1+n) z_n,
\end{eqnarray*}
where we have used identities (\ref{sum1}) and (\ref{sum2}).  Therefore, we obtain
\begin{eqnarray}
 \langle \sigma_{\theta}^2 \rangle = \frac{2 \pi^2}{3 S^2} \left[ \left( 1 - S z_0 \right) (S-1) (2S-1) -  
    2 \sum_{n=1}^{S-2} (S-1-n)(S-n)(2S-1+n) z_n  \right].    
    \label{sigma-ave-4}
\end{eqnarray}
Using the approximate expression for $z_0$ from Eq.~(\ref{zn2}) we obtain, for the $S=2$ case, 
\begin{eqnarray*}
	\langle \sigma_{\theta}^2 \rangle = \frac{\pi^2}{2} (1-2 z_0) \simeq \frac{\pi^2}{4} \left( 1- \frac{1}{N} \right),
\end{eqnarray*}
which agrees with Eq.~(\ref{sigma-ave-2}) obtained by direct integration (see section~\ref{S2}).  For the case $S \ge 3$, we replace the moments $z_n$ in Eq.~(\ref{sigma-ave-4}) by the approximate expressions from Eq.~(\ref{zn3}) and arrive to 
\begin{eqnarray}
  \label{sigma-ave-5}
   \langle \sigma_{\theta}^2 \rangle = \frac{2 \pi^2}{3 S^2} \left[ \left( 1 - \frac{1}{S} 
    + \frac{1-S^2}{4SN}  \right) (S-1)(2S-1) -  \frac{2}{S^2} \left( 1 - \frac{1-S^2}{4N} \right) c_S  + 
    \frac{3 \, d_S}{2 S^2 N} \right],
\end{eqnarray}
with
\begin{eqnarray*}
  c_S &=& \sum_{n=1}^{S-2} (S-1-n)(S-n)(2S-1+n) ~~~\mbox{and} \\
  d_S &=& \sum_{n=1}^{S-2} (S-1-n)(S-n)^2 (2S-1+n) n.
\end{eqnarray*}
To calculate the coefficients $c_S$ and $d_S$ above we expand the terms of each summation in powers of $n$ and use the identities Eqs.~(\ref{sum1}-\ref{sum5}) to obtain, after some algebra, 
\begin{eqnarray*}
    c_S &=& \frac{S(S-1)(S-2)(3S-1)}{4} ~~~\mbox{and} \\
    d_S &=& \frac{S^2(S^2-1)(S-2)(7S-1)}{60}.
\end{eqnarray*}
Replacing the above expressions for $c_S$ and $d_S$ in Eq.~(\ref{sigma-ave-5}) and simplifying the resulting expression we finally arrive to
\begin{eqnarray}
\langle \sigma_{\theta}^2 \rangle \simeq \frac{\pi^2 (S^2-1)}{3 S^2} \left( 1 - \frac{S^2+11}{20N} \right) ~~~ \mbox{for $\Delta=1$, $S \ge 3$ and $N \gg S^2$.} 
  \label{sigma-ave-6}
\end{eqnarray}
Equation~(\ref{sigma-ave-6}) tells that the average width of the angular states distribution is smaller than that of the uniform distribution $\sigma_u$, and that approaches $\sigma_u^2=\pi^2 (S^2-1)/(3 S^2)$ as $1/N$ when $N$ increases.  As previously suggested, this result shows that the distribution of angular states becomes uniform in the $N \to \infty$ limit, where disorder is total ($\psi=0$).

\begin{figure}[t]
  \centerline{\includegraphics[width=8.0cm]{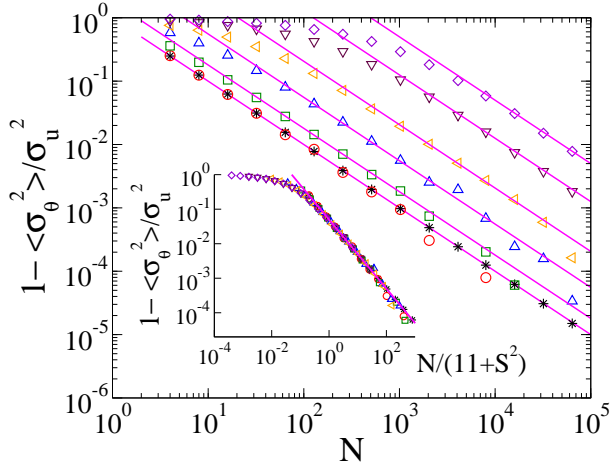}}
  \caption{Average value of the mean-squared deviation $\sigma_{\theta}^2$ at the stationary state vs system size $N$ for error amplitude $\Delta=1$ and number of angular states $S=2$ (stars), $S=3$ (circles), $S=5$ (squares), $S=10$ (up triangles), $S=20$ (left triangles), $S=50$ (down triangles) and $S=100$ (diamonds).  Straight lines are the theoretical approximation given by Eq.~(\ref{sigma-ave-6}).  The inset shows the collapse of the data when the $x$-axis is rescaled by $11+S^2$, and that the analytical approximation from Eq.~(\ref{sigma-ave-6}) (straight line) is already valid for $N \gtrsim S^2$.}
  \label{sigma-N-delta1}
\end{figure}

Figure~\ref{sigma-N-delta1} shows MC simulation results for the behavior of $\langle \sigma_{\theta}^2 \rangle$ with $N$, for $\Delta=1$ and various values of $S$ (symbols).  We see that the agreement with Eq.~(\ref{sigma-ave-6}) (solid lines) is good for $N \gtrsim S^2$, as it happens with $\langle \psi \rangle$.  This approximate lower limit for the validity of Eq.~(\ref{sigma-ave-6}) can be better checked in the data collapse shown in the inset.

\begin{figure}[t]
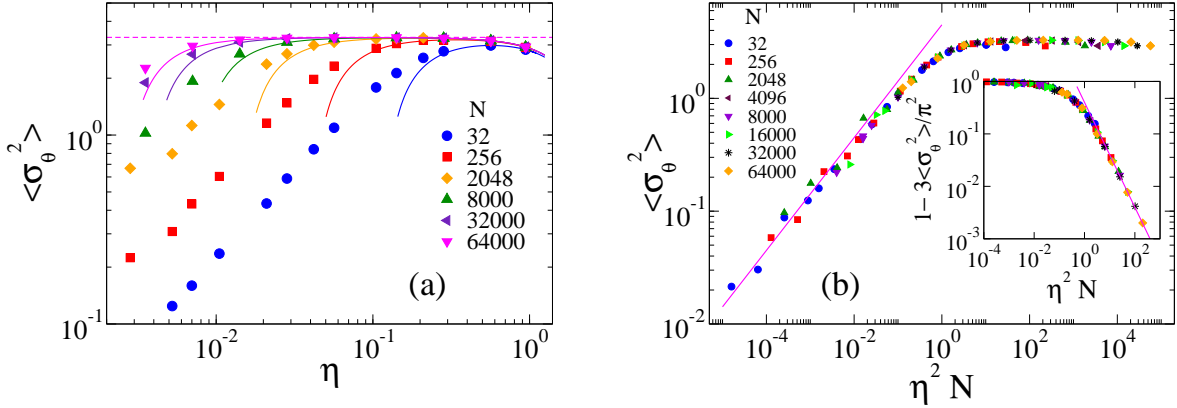

  \begin{center}
    \vspace{0.5cm}
    \begin{tabular}{cc}
      \hspace{-1.0cm}
      \includegraphics[width=5.0cm, bb=70 -20 550 550]{Fig8a.eps}
      & \hspace{3.0cm} 
      \includegraphics[width=5.0cm, bb=70 -20 550 550]{Fig8b.eps} 
    \end{tabular}    
    \caption{(a)  Average value of $\sigma_{\theta}^2$ vs relative error amplitude $\eta(\Delta,S)$ for the system sizes $N$ indicated in the legend.  The values of $\Delta$ and $S$ are the same as those of Fig.~\ref{psi-eta-all}.  Solid curves are the analytical approximation Eq.~(\ref{sigma-ave-7}) for $\eta \gg N^{-1/2}$.  The horizontal dashed line is the saturation value $\sigma_u^2=\pi^2/3$ obtained for $\eta \ll 1$ in the
$N \to \infty$ limit.  (b)  The data collapses into a single curve when it is plotted vs the rescaled variable $\eta^2 N$.  The main plot shows the behavior of $\langle \sigma_{\theta}^2 \rangle$ for very small values of $\eta^2 N$.  The straight line has slope $1/2$.  The inset shows how $\langle \sigma_{\theta}^2 \rangle$ approaches the value $\sigma_u^2$ as $N$ increases, when $\eta$ is small (see main text).  The solid line is the approximation $1- \langle \sigma_{\theta}^2 \rangle/\sigma_{u}^2 \simeq 2/(5 \, \eta^2 N)$ from Eq.~(\ref{sigma-ave-7}) for small $\eta$.}
    \label{sigma-eta-all}
  \end{center}
\end{figure}

To generalize Eq.~(\ref{sigma-ave-6}) for any noise amplitude $\eta$, we follow an analysis similar to the one done in section~\ref{section-psi} for $\langle \psi \rangle$.  Replacing the expression $S=2\sqrt{2}/\eta$ obtained for $\Delta=1$ into Eq.~(\ref{sigma-ave-6}) we obtain 
\begin{eqnarray}
  \langle \sigma_{\theta}^2 \rangle \simeq \frac{\pi^2(8-\eta^2)}{24} \left( 1 - \frac{8+11 \, \eta^2}{20 \, \eta^2 N} \right) ~~~~~ \mbox{for $S \ge 3$ and
    $N \gg 1/\eta^2$,} 
  \label{sigma-ave-7}
\end{eqnarray}
with $\eta=\eta(\Delta,S)$ given by Eq.~(\ref{eta-Delta-S}).  To test Eq.~(\ref{sigma-ave-7}) we performed MC simulations for various system sizes and various different values of the set $(\Delta,S)$, and calculated the average mean-squared deviation.  Results are shown by symbols in Fig.~\ref{sigma-eta-all}.  In panel (a) we plot $\langle \sigma_{\theta}^2 \rangle$ as a function of $\eta(\Delta,S)$.  We see that $\langle \sigma_{\theta}^2 \rangle$ increases with $\eta$ and saturates at the value $\sigma_u^2 \simeq \pi^2/3$ corresponding to the large number of states ($S>10$) used in simulations, and then decreases for larger $\eta$.  We also observe that, for a given $N$, Eq.~(\ref{sigma-ave-7}) (solid lines) gives a good estimation of the numerical data for the largest values of $\eta$.
The collapse of the data in panel (b) shows that $\langle \sigma_{\theta}^2 \rangle$ is a function of $\eta^2 N$, as long as $\eta$ is small enough.  This behavior is in agreement with Eq.~(\ref{sigma-ave-7}), from where we see that for small $\eta$ is $\langle \sigma_{\theta}^2 \rangle \simeq \frac{\pi^2}{3} \left( 1 - \frac{2}{5 \, \eta^2 N} \right)$.  This theoretical approximation is plotted by a solid line in the inset, showing the power-law approach $2/(5 \, \eta^2 N)$ of $\langle \sigma_{\theta}^2 \rangle$ to $\pi^2/3$ that is valid when $\eta^2 N \gtrsim 1$.  For the sake of clarity, data points for large values of $\eta$ that fall off the straight line were removed.

We can also see in the main plot of Fig.~\ref{sigma-eta-all}(b) that for $\eta^2 N \lesssim 10^{-2}$ the data seems to follow a power-law increase with an exponent similar to $1/2$, suggesting the scaling $\langle \sigma_{\theta}^2 \rangle \sim \eta N^{1/2}$ (solid line).  In the next section we give a  theoretical insight into this particular behavior of the system in the low noise limit.

\section{The continuum approach for large $S$} 
\label{continuum}

In this section we consider the limiting case $S \gg \Delta \gg 1$, that is, the limit of a large number of particles' states and a very small amplitude of the relative error, $\eta \simeq 2 \Delta/S \ll 1$ [see Eq.~(\ref{eta-Delta-S})].  Before entering into the definition of the dynamics, we describe bellow a series of assumptions that we make to simplify our analysis.

In the limit of very large $S$ the angular space becomes continuous, and thus the state of a given particle $i$ at time $t$ can be taken as a real variable $\Theta_i(t)$ in the continuous space.  The noise $\xi_i(t)$ introduced in the state of particle $i$ after the copying process at time $t$ can also be considered as a continuous variable uniformly distributed in the interval $[-\eta \pi,\eta \pi]$ (uniform white noise), with first moment $\langle \xi_i(t) \rangle=0$ and second moment $\langle \xi_i^2(t) \rangle=\pi^2 \eta^2/3$ for all $i$ and $t$.  Moreover, when the noise amplitude $\eta$ is very small we expect that the distribution of particles' states will be very narrow at the stationary state, as compared to the length $2 \pi$ of the angular space (see for instance down-left panel of Fig.~\ref{xk-2}).  Thus, all particles will be far from the borders at $0$ and $2 \pi$ during the time we consider in this analysis.  Therefore, we assume that particles can freely diffuse in the entire real axis with no boundary conditions.  Besides, we consider that the updates of particles' states take place in parallel (all at the same time) at discrete integer times $t=0,1,2...$, which resembles the update of the Wright-Fisher model in contrast to the sequential update of the Moran or voter dynamics implemented in our model.  As each particle interacts once in average per unit time, we expect that both the \emph{sequential} and the \emph{parallel} updates have similar mesoscopic behavior.  In fact, it was shown in \cite{Blythe-2007} that the Wright-Fisher and Moran models give the same mesoscopic Fokker Planck equation.

Let us consider that at a given time $t-1$ the states of particles are described by the angles $\Theta_i(t-1)$, with $i=1,..,N$.  Then, in a single step of the parallel dynamics, for each particle $i$ we select a random particle $j \ne i$ and update its state according to $\Theta_i(t-1) \to \Theta_i(t)=\Theta_j(t-1)+\xi_i(t)$, where $\xi_i(t) \in [-\eta \pi, \eta \pi]$ is a small perturbation in the form of noise.  We remark that the state of every particle at time $t$ depends on the state of another particle at the previous time $t-1$, and that time is increased by $\Delta t=1$ after all particles have updated their states.  Suppose that at time $t-1$ particle $i$ copies the state of particle $j$, who has copied the state of particle $k$ at the previous time $t-2$ with a perturbation $\xi_j(t-1)$.  Then, the state of $i$ at time $t$ can be expressed as $\Theta_i(t)=\Theta_k(t-2)+\xi_j(t-1)+\xi_i(t)$ and, iterating back in time until $t=0$, as
\begin{equation}
  \Theta_i(t) = \sum_{\tau=0}^t \delta_i^{\tau}(t),
  \label{sup}
\end{equation}
where $\delta_i^0(t)=\Theta_m(0)$ is the initial state of some particle $m$,
$\delta_i^{\tau}(t)=\xi_n(\tau)$ (for $1 \le \tau \le t-1$) corresponds to the noise added to some particle $n$ at time $\tau$, and $\delta_i^t(t)=\xi_i(t)$.  The reason why we use the notation $\delta_i^{\tau}(t)$ is because each term in the summation of Eq.~(\ref{sup}) depends on the present time $t$ and the past time $\tau$, as we shall see bellow.  Given that the set of numbers $\delta_i^\tau(\tau)=\xi_i(\tau)$ introduced in the system at $t=\tau$ will play a special role in the rest of this section, we will refer to them with the name of \textit{generation} $\tau$.

\begin{figure}[t]
  \centerline{\includegraphics[width=10.0cm]{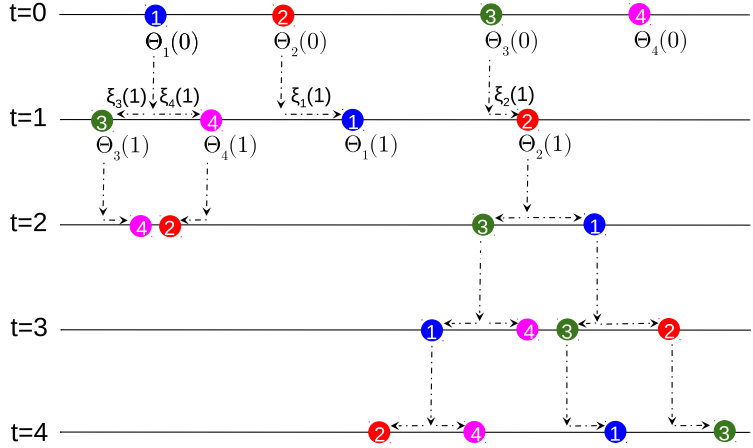}}
  \caption {This illustration shows the evolution of a system with four particles in a single realization, as described in the text.  Time runs vertically downward.  The state of all particles are updated in parallel at each time step of the dynamics.  Vertical dashed arrows indicate the update of a particle's state after copying the state of another particle, while the horizontal dashed arrows represent the effect of the perturbation by noise.}
  \label{time-superpose}
\end{figure}

To help better understand the dynamics of the system we show in Fig.~\ref{time-superpose} a simple example for the evolution of a four-particle system in a single realization.  Initially, particles have states $\Theta_i(0)$ ($i=1,2,3,4$).  In the first time step, particle $1$ copies the state of particle $2$, $2$ copies the state of $3$, while $3$ and $4$ copy the state of $1$, and then a different noise $\xi_i(1)$ is added to each particle.  Thus, particles' states at $t=1$ can be written as
\begin{eqnarray*}
  \Theta_1(1) &=& \Theta_2(0) + \xi_1(1), \\
  \Theta_2(1) &=& \Theta_3(0) + \xi_2(1), \\
  \Theta_3(1) &=& \Theta_1(0) + \xi_3(1), \\
  \Theta_4(1) &=& \Theta_1(0) + \xi_4(1).   
  \label{ex1}  
\end{eqnarray*}

In terms of generations and by looking at the right-hand side of these equations, we can say that at time $t=1$ only $3$ different states are alive in generation $\tau = 0$ [$\Theta_1(0)$, $\Theta_2(0)$ and $\Theta_3(0)$], while generation $\tau=1$ has just been born with its $4$ different states [$\xi_1(1)$, $\xi_2(1)$, $\xi_3(1)$ and $\xi_4(1)$].  The terms in the summation Eq.~(\ref{sup}) for particle $1$ are $\delta_1^0(1)=\Theta_2(0)$ and $\delta_1^1(1)=\xi_1(1)$, and similarly for the states of the other three particles.  In the second time step $t=2$, $1$ and $3$ copy $2$, $2$ copies $4$, and $4$ copies $3$.  Then, $\Theta_1(2)=\Theta_2(1)+\xi_1(2)=\Theta_3(0)+\xi_2(1)+\xi_1(2)$ and thus $\delta_1^0(2)=\Theta_3(0), \delta_1^1(2)=\xi_2(1)$ and $\delta_1^2(2)=\xi_1(2)$.
We note that the dependence in $t$ of each term $\delta_1^{\tau}(t)$ is due to the fact that when particle $1$ copies a particle $j$ it replaces all $t-1$ terms in the summation by those of $j$ at $t-1$.  This process can be seen as particle $1$ copying the ``state history'' of particle $j$ from time $0$ to $t-1$, formed by the list of $t-1$ terms in the summation $\Theta_j(t-1)=\sum_{\tau=0}^{t-1} \delta_j^{\tau}(t-1)$.  In a similar way, we can iterate the copying processes at each time step as depicted in Fig.~\ref{time-superpose}, and find the following states at $t=4$:
\begin{eqnarray}
  \Theta_1(4) &=& \Theta_3(0) + \xi_2(1) + \xi_1(2) + \xi_3(3) + \xi_1(4), \nonumber \\
  \Theta_2(4) &=& \Theta_3(0) + \xi_2(1) + \xi_3(2) + \xi_1(3) + \xi_2(4), 
\nonumber \\
  \Theta_3(4) &=& \Theta_3(0) + \xi_2(1) + \xi_1(2) + \xi_2(3) + \xi_3(4), \nonumber \\
  \Theta_4(4) &=& \Theta_3(0) + \xi_2(1) + \xi_3(2) + \xi_1(3) + \xi_4(4).  
  \label{ex2}  
\end{eqnarray}
It is interesting to note in Eqs.~(\ref{ex2}) that, even though all four particles started from different states $\Theta_i(0)$ at $t=0$ [$\delta_i^0(0)=\Theta_i(0)$], the state history of all particles at $t=4$ have the same ``parent'' particle $3$ (root) at $\tau=0$ [$\delta_i^0(4)=\Theta_3(0) ~ \forall i$].  That is, generation $\tau=0$ has already converged to $\Theta_3(0)$ at time $t=4$.  They also have the same parent particle $2$ at $\tau=1$ [$\delta_i^1(4)=\xi_2(1) ~ \forall i$], i e., generation $\tau=1$ converged to $\xi_2(1)$ at $t=4$.  We can also see that in  Fig.~\ref{time-superpose}, where the states of all particles at $t=4$ can be traced back in time to the state of particle $2$ at $t=1$, and to the state of particle $3$ at $t=0$.  That is, when we look backwards in time, this ``tree structure'' that emerges can be seen as a system of coalescing random walks, which is the dual process of the voter model, as it is known in the mathematics literature on voter models \cite{Cox-1989}.  In general, if we look at a given time $\tau$, all four different numbers $\delta_i^\tau(\tau)=\xi_i(\tau)$ introduced at $t=\tau$ (generation $\tau$) change at each time step $t > \tau$ to the new values $\delta_i^{\tau}(t)$ following the copying dynamics between the numbers $\delta_i^{\tau}$ of the same generation $\tau$.  Therefore, we can see $\delta_i^{\tau}$ as the state of the particle $i$ in generation $\tau$ that evolve under the rules of the original MSVM (without noise).  In general terms, each generation $\tau$ behaves as a MSVM without noise that starts with $N$ different random states $\delta_i^{\tau}(\tau)=\xi_i(\tau)$ at time $\tau>0$, or from $\delta_i^0(0)=\Theta_i(0)$ at $\tau=0$.  Given that convergence into a single state is eventually reached in the voter dynamics, we expect that the number of different states $s_{\tau}$ of a given generation $\tau$ decreases with time until consensus ($s_{\tau}=1$) is achieved when the generation has evolved for a time $t-\tau \sim N$, which happens in the example of Fig.~\ref{time-superpose} for $t=4$ and $\tau=0$ and $\tau=1$.  These observations will be relevant for the calculations ahead.  Contrary to the order of section~\ref{S3}, we first derive a scaling behavior for $\sigma_{\theta}^2$ that we use then to estimate $\psi$.

\subsection{Calculation of the mean-squared deviation $\langle \sigma_{\theta}^2 \rangle$}
\label{section-sigma-2}

Now that we have described basic properties of the dynamics, let us start the analysis of the system by calculating the mean-squared deviation (sample variance) $\sigma_{\theta}^2(t)=\frac{1}{N} \sum_{i=1}^N \left[ \Theta_i(t)-\overline{\theta}(t)\right]^2$, which can be expressed as
\begin{equation}
  \sigma_{\theta}^2(t)=\frac{1}{N} \sum_{i=1}^N \left[ \sum_{\tau=0}^t \Delta \delta_i^{\tau}(t) \right]^2,
  \label{sigsup}
\end{equation}
where we have used Eq.~(\ref{sup}) and introduced the new variable
\begin{equation}
  \Delta \delta_i^{\tau}(t) \equiv \delta_i^{\tau}(t)-\overline{\delta^{\tau}}(t), ~~~ \mbox{with ~~ $\overline{\delta^{\tau}}(t)=\frac{1}{N} \sum_{i=1}^N \delta_i^{\tau}(t)$}.
  \label{Delta}
\end{equation}
Then, the expectation value of $\sigma_{\theta}^2(t)$ over different runs of the dynamics can be written from Eq.~(\ref{sigsup}) as
\begin{equation}
  \langle \sigma_{\theta}^2(t) \rangle = \frac{1}{N} \sum_{i=1}^N \sum_{\tau=0}^t \langle \left[ \Delta \delta_i^{\tau}(t) \right]^2 \rangle + \frac{2}{N} \sum_{i=1}^N \sum_{\tau=0}^{t-1} \sum_{\tau > \tau'}^t \langle \Delta \delta_i^{\tau}(t) \, \Delta \delta_i^{\tau'}(t) \rangle. 
  \label{sigsup-2}
\end{equation}
Given that the perturbations introduced in a given generation $\tau$ are independent of the perturbations introduced in a different generation $\tau' \neq \tau$, we have $\langle \mathcal \delta_i^{\tau}(t) \, \delta_j^{\tau'}(t)\rangle=\langle \xi_n(\tau) \, \xi_m(\tau') \rangle=\langle \xi_n(\tau) \rangle \langle \xi_m(\tau') \rangle=0$ for all $i$ and $j$.  Using this last relationship and Eq.~(\ref{Delta}) one can check that $\langle \Delta \delta_i^{\tau}(t) \, \Delta \delta_i^{\tau'}(t) \rangle=0$ for $\tau' \neq \tau$, and thus the second term of Eq.~(\ref{sigsup-2}) vanishes, leading to the simple expression
\begin{equation}
  \langle \sigma_{\theta}^2(t)\rangle=\sum_{\tau=0}^t \langle \sigma_{\tau}^2(t)\rangle.
  \label{sigsup-3}
\end{equation}
Here $\sigma_{\tau}^2(t) \equiv \frac{1}{N} \sum_{i=1}^N \left[ \Delta \delta_i^{\tau}(t) \right]^2$ is the variance at time $t$ of the perturbations $\delta_i^{\tau}(\tau)$ introduced at time $\tau$ (we have used $\sigma_{\tau}=\sigma_{\delta^{\tau}}$ to simplify notation).  We see that the expectation value of the dispersion in the particles' states at time $t$ can be expressed as the superposition of the corresponding expectation values in the generations introduced in the previous times $0 \le \tau \le t$, which have evolved under the multi-state voter dynamics during a time $t-\tau$.

We can now use the basic known results of the MSVM described in section~\ref{model} to study the evolution of the variance in a given generation $\tau$.  Starting from $N$ states $\delta_i^{\tau}(\tau)$ at time $\tau$, the number of different states occupied by particles in generation $\tau$,  $s_{\tau}$, decreases with time due to the copying process until all particles condensate into one state.  Therefore, we expect that the dispersion will decrease with time and reach the value $\sigma_{\tau}^2=0$ when consensus in that generation is reached.  From Eq.~(\ref{s-t}), the mean number of states in generation $\tau$ behaves approximately as 
\begin{eqnarray}
  s_{\tau}(t) \simeq
  \begin{cases}
    \frac{N}{1+(t-\tau)/2} & \mbox{for $0 \le t-\tau \lesssim N$}, \\
    1 & \text{for $t-\tau \gtrsim N$},
  \end{cases}
  \label{St}
\end{eqnarray}
where we have assumed an initial value $s_{\tau}(\tau)=S=N \gg 1$ and introduced the prefactor $2$ that appears in the parallel update (see \cite{Baglietto-2018}).  Note also that $s_{\tau}$ is $1$ after consensus is reached in a time of order $N$.  Then, if at time $t$ there are $s_{\tau}(t) \le N$ surviving states that we call $\tilde \delta_k^{\tau}(t)$ [$k=1,..,s_{\tau}(t)$], we can express the states' variance as $\sigma_{\tau}^2(t) \simeq \frac{1}{N} \sum_{k=1}^{s_{\tau}(t)} n_k(t) \left[ \tilde \delta_k^{\tau}(t) - \overline{\tilde \delta^{\tau}}(t) \right]^2$, where $n_k$ is the number of particles with state $\tilde \delta_k^{\tau}$ and $\overline{\tilde \delta^{\tau}}=\frac{1}{s_{\tau}} \sum_{k=1}^{s_{\tau}} \tilde \delta_k^{\tau}$ is their mean value (note that $k$ is the state label rather than the particle label).  Assuming that the $N$ particles are distributed uniformly among the $s_{\tau}(t)$ states, we can write  $n_k(t) \simeq N/s_{\tau}(t)$ $\forall k$ and thus
\begin{equation}
  \sigma_{\tau}^2(t) \simeq \frac{1}{s_{\tau}(t)} \sum_{k=1}^{s_{\tau}(t)} \left[ \tilde \delta_k^{\tau}(t) - \overline{\tilde \delta^{\tau}}(t) \right]^2.
  \label{sigsup-4}
\end{equation}
Expanding the right-hand side of Eq.~(\ref{sigsup-4}) we find the products $\tilde \delta_k^{\tau} \, \tilde \delta_{k'}^{\tau}$, whose expectation value is
\begin{equation}
  \langle \tilde \delta_k^{\tau}(t) \, \tilde \delta_k^{\tau'}(t) \rangle = \frac{\pi^2 \eta^2}{3} \, \delta_{k,k'},
  \label{deltas}
\end{equation}
where now $\delta_{k,k'}$ is the Kronecker delta.  Applying brackets to both sides of Eq.~(\ref{sigsup-4}) and using relations Eq.~(\ref{deltas}) we obtain
\begin{equation}
  \langle \sigma_{\tau}^2(t) \rangle \simeq \frac{\pi^2 \eta^2}{3} \frac{\left[ s_{\tau}(t)-1 \right]}{s_{\tau}(t)},  
  \label{sigsup-5}
\end{equation}
which finally becomes
\begin{eqnarray}
  \langle \sigma_{\tau}^2(t) \rangle \simeq
  \begin{cases}
    0 & \text{for $0 \le \tau \lesssim t-N$}, \\
    \frac{\pi^2 \eta^2}{6N}\left( 2N-t+\tau \right) & \mbox{for $t-N \lesssim \tau \le t$},
  \end{cases}
  \label{sigsup-6}
\end{eqnarray}
after replacing expression Eq.~(\ref{St}) for $s_{\tau}$.  Plugging Eq.~(\ref{sigsup-6}) into Eq.~(\ref{sigsup-3}) and performing the summation 
\begin{equation}
  \langle \sigma_{\theta}^2(t)\rangle \simeq \frac{\pi^2 \eta^2}{6N} \sum_{\tau=t-N}^t (2N-t-\tau)
  \label{sigsup-7}
\end{equation}
we finally arrive to 
\begin{equation}
  \langle \sigma_{\theta}^2(t)\rangle \simeq c \, \eta^2 N ~~~\mbox{for $t \ge N$},
  \label{sigsup-8}
\end{equation}
where $c$ is a constant.  It is interesting to note from Eq.~(\ref{sigsup-7}) that for any time $t \ge N$, on average, only the $N$ generations that are less than a distance $N$ from $t$ contribute to $\langle \sigma_{\theta}^2(t)\rangle$, given that generations earlier than $t-N$ have reached consensus and thus they have zero variance.  This makes $\langle \sigma_{\theta}^2 \rangle$ reach a stationary value when $t \ge N$.

\begin{figure}[t]
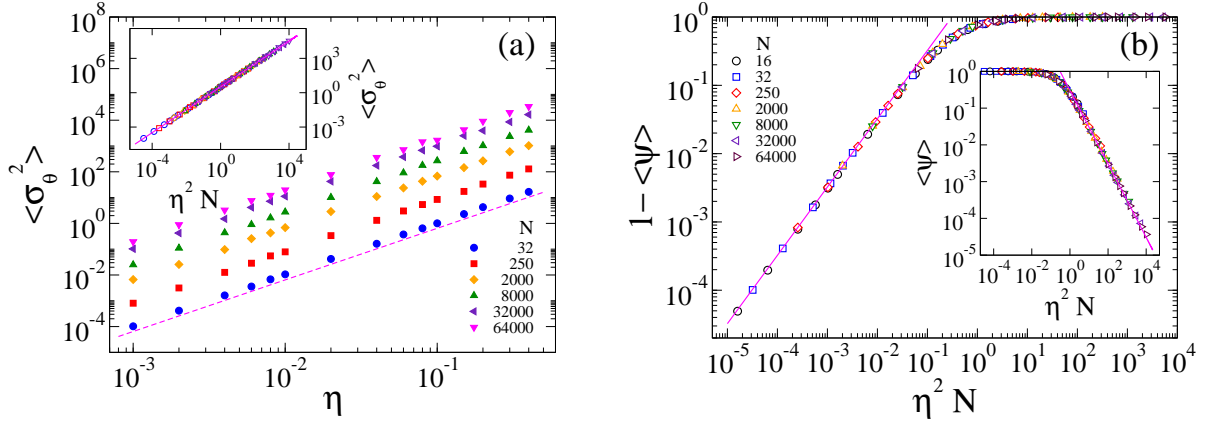

  \begin{center}
    \vspace{0.5cm}
    \begin{tabular}{cc}
      \hspace{-1.0cm}
      \includegraphics[width=5.0cm, bb=70 -20 550 550]{Fig10a.eps}
      & \hspace{3.0cm} 
      \includegraphics[width=5.0cm, bb=70 -20 550 550]{Fig10b.eps} 
    \end{tabular}    
    \caption {Simulation results of the model under the parallel update for continuous states and noise, and for the systems sizes indicated in the legends.  a) $\langle \sigma_{\theta}^2 \rangle$ vs $\eta$ on a double-logarithmic scale.  The dashed line has slope $2$.  Inset: Data collapse.  The straight line is the analytical approximation $c \, \eta^2 N$ [Eq.~(\ref{sigsup-8})], with $c=3.24$ corresponding to the best fit.  (b) The collapse of the data shows that the order parameter $\langle \psi \rangle$ approaches $1$ as $1 - c \, \eta^2 N$ when $\eta^2 N \to 0$ (straight line).  Inset: $\langle \psi \rangle$ vanishes as $3/(\pi^2 \eta^2 N)$ when $N \to \infty$ (straight line).}
    \label{sigma-eta-all-para}
    \end{center}    
\end{figure}

In order to test the previous result Eq.~(\ref{sigsup-8}), we have run simulations of the model in continuous state space (states $\theta$ and noise $\xi$ are real numbers in $[-\infty,\infty]$ and $[-\eta,\eta]$, respectively) under the parallel update.  Fig.~\ref{sigma-eta-all-para}(a) shows that, for various system sizes, $\sigma_{\theta}^2$ grows with $\eta$ as a power law with exponent $2$ (dashed lines).  Also, the data collapse in the inset confirms the scaling given by Eq.~(\ref{sigsup-8}), where the straight line is the function $c \, \eta^2 N$, with $c=3.24$ corresponding to the best fit (solid line).

Eq.~(\ref{sigsup-8}) predicts that the distribution of angular states at the stationary state has a width that increases linearly with $\eta^2$ and $N$ when the system is unbounded, given that particles' states can freely spread on the real axis.  However, if the system has periodic boundaries at $\theta=0$ and $\theta=2 \pi$, particles are bounded in $[0,2 \pi)$ and thus the width saturates to the value $\sigma_u \simeq \pi/\sqrt{3}$ when $N$ and $\eta$ increase, as explained in section~\ref{section-sigma}.  Besides that this saturation is obviously not captured by the open boundary approach developed in this section, neither the scaling $\langle \sigma_{\theta}^2 \rangle \sim \eta N^{1/2}$ observed in Fig.~\ref{sigma-eta-all}(b) for small $\eta^2 N$ (solid line) agrees with that of Eq.~(\ref{sigsup-8}).  The reason for that discrepancy lies in the fact that, for a single realization in the periodic system, all particles concentrate in a narrow interval whose distribution has a mean-squared deviation that scales as $\langle \sigma_{\theta}^2 \rangle \sim \eta^2 N$ as long as particles are far from the boundaries (not shown), but when particles reach a boundary the distribution splits into two separate sharp distributions peaked at $0$ and $2\pi$, increasing its mean-squared deviation from the small value $\sigma_{\theta}^2 \sim \eta^2 N$ to a much larger value $\sigma_{\theta}^2 \simeq 2 \pi^2$.  This last value of $\sigma_{\theta}^2$ is roughly estimated assuming that particles are concentrated around the boundaries and thus the distribution looks like two Dirac delta functions at $0$ and $2 \pi$.  Then, when 
$\sigma_{\theta}^2$ is measured at a given time of a single realization, we estimate that the probability $\omega$ that the distribution is split into two parts is proportional to the distribution's width.  That is, with probability $\omega \propto \sigma_{\theta}/2 \pi \propto \eta N^{1/2}$ is $\sigma_{\theta}^2 \simeq 2 \pi^2$ and with the complementary probability $1-\omega$ is $\sigma_{\theta}^2 \propto \eta^2 N$.  This leads to an average mean-squared deviation over many runs that scales as 
\begin{equation}
  \langle \sigma_{\theta}^2 \rangle \sim \omega \, 2 \pi^2 + (1-\omega) \eta^2 N \sim \eta N^{1/2} + \mathcal O(\eta^2 N) ~~~ \mbox{for $\eta N^{1/2} \ll 1$}.
  \label{sigsup-9}
\end{equation}
Equation~(\ref{sigsup-9}) gives the correct scaling $\langle \sigma_{\theta}^2 \rangle \sim \eta N^{1/2}$ obtained in the simulations for the discrete system under the sequential update [Fig.~\ref{sigma-eta-all}(b)] when $\eta^2 N \lesssim 10^{-2}$ (solid line with slope $1/2$).

\subsection{Calculation of the order parameter $\langle \psi \rangle$}
\label{section-psi-2}

For the low noise case, we shall see bellow that there is a simple relationship between the order parameter 
$\psi(t)$ and the variance $\sigma_{\theta}^2(t)$ of the particle system, which allows to estimate $\langle \psi \rangle$ from the behavior of $\langle \sigma_{\theta}^2 \rangle$ found in the last subsection.  

As we showed before, when the noise is very low the states of all particles are within a very narrow angular window that, for the sake of simplicity, it is assumed to be centered at $\theta=0$ given that $\psi$ is invariant under angular translations.  Indeed, one can check from the definition Eq.~(\ref{psi}) of the order parameter that any translation $\Theta_m \to \Theta_m+\alpha$ ($m=1,..,N$) returns the same value of $\psi$.  Therefore, we can approximate the exponential functions in Eq.~(\ref{psi}) as $e^{i \Theta_m} \simeq  1 + i \Theta_m - \Theta_m^2/2$ to second order in $|\Theta_m| \ll 1$, and write the order parameter as 
\begin{equation}
 \psi(t)  \simeq \left| 1 + i \, \overline{\theta}(t) - \frac{1}{2} \overline{\theta^2}(t) \right|^2 = 1 - \overline{\theta^2}(t) + \overline{\theta}^2(t) + \mathcal O(\theta^4) = 1 - \sigma_{\theta}^2(t) + \mathcal O(\theta^4), 
\label{psi-low}
\end{equation}
where $\overline{\theta}$ and $\overline{\theta^2}$ are the first and second moments of the states' distribution defined in Eqs.~(\ref{moments-theta}), and $\sigma_{\theta}^2=\overline{\theta^2} - \overline{\theta}^2$.  From Eq.~(\ref{psi-low}), the average value of the order parameter at the stationary state is related to the average mean-squared deviation of the population by the simple expression
\begin{equation} 
  \langle \psi \rangle \simeq 1 - \langle \sigma_{\theta}^2 \rangle.
  \label{psi-low-2}
\end{equation}
As expected, the order increases as the width of the distribution of particles decreases, and reaches its maximum value $\psi=1$ at consensus ($\sigma_{\theta}^2=0$).  Finally, replacing Eq.~(\ref{sigsup-8}) for 
$\langle \sigma_{\theta}^2 \rangle$ into Eq.~(\ref{psi-low-2}) we obtain
\begin{equation} 
  \langle \psi \rangle \simeq 1 - c \, \eta^2 N.
  \label{psi-low-3}
\end{equation}
Figure~\ref{sigma-eta-all-para}(b) shows the behavior of the order parameter $\langle \psi \rangle$ with the rescaled variable $\eta^2 N$, obtained from simulations of the continuous model under the parallel update for various system sizes.  We observe that for $\eta^2 N \lesssim 10^{-2}$ the data is in good agreement with Eq.~(\ref{psi-low-3}) (solid line) using the value $c=3.24$ obtained from the best fit of the $\langle \sigma_{\theta}^2 \rangle$ vs $\eta^2 N$ data shown in the inset of Fig.~\ref{sigma-eta-all-para}(a).  In the inset of Fig.~\ref{psi-eta-all}(b) we show that Eq.~(\ref{psi-low-3}) with a constant $c_s=1.64$ (solid line) describes the behavior of $\langle \psi \rangle$ at low noise for the discrete sequential version of the model as well.  The prefactor similar to $1/2$ between the constants $c_s$ and $c$ of the two versions of the model also appears in the expression $3/(\pi^2 \eta^2 N)$, as compared to Eq.~(\ref{psi-6}), which reproduces very well the behavior of $\langle \psi \rangle$ for large $N$ of the continuous parallel model shown in the inset of Fig.~\ref{sigma-eta-all-para}(b).  We do not know how to explain this prefactor.     

Equation~(\ref{psi-low-3}) shows that for a system of size $N$, total order $\langle \psi \rangle=1$ is eventually achieved as the noise vanishes.  This result completes the picture of the behavior of the model in the two limits.  That is, for fixed $\eta>0$, complete disorder is reached in the $N \to \infty$ limit, while for fixed $N>0$, complete order is achieved as $\eta \to 0$.

\section{Summary and Conclusions}
\label{conclusions}

We studied the dynamics of a multi-state voter model in mean-field with a degree of error or imperfection in the copying process.  Starting from a uniform distribution of particles over a discrete angular space, we investigated the dynamics of ordering and its stationary state.  When the copying is perfect the number of different angular states occupied by particles decreases monotonically with time until eventually the system reaches an absorbing state of complete order where all particles share the same state.  However, when we add a source of imperfection in the copying process that leaves the states of two interacting particles similar but not exactly equal (an imperfect copying) a new scenario appears.  The system evolves towards a stationary state characterized by an ordering level $\psi$ that depends on the number of particles $N$, the number of possible particle states $S$ and the copying error amplitude $\Delta$.  We analyzed two different limits.  In the large $N$ limit we proved by means of a Fokker-Planck equation approach that the average order decreases with $N$ and the relative error amplitude $\eta = 2 \Delta/S$ as $\langle \psi \rangle \simeq 6/(\pi^2 \eta^2 N)$ for $0 < \eta \ll 1$ and $\eta^2 N \gtrsim 1$.  Besides, when $N$ and $\eta$ increase, the average mean-squared deviation of particles' states approaches the value $\sigma_u^2$ corresponding to the uniform distribution as $\langle \sigma_{\theta}^2 \rangle \simeq \sigma_u^2 [1-2/(5\, \eta^2 N)]$.  These results imply that for any degree of error $\eta>0$ the system gets completely disordered in the thermodynamic limit $N \to \infty$, where the distribution of particles over the angular space is perfectly uniform.  In the large $S$ limit we developed an analytical approach that assumes a continuum angular space and showed that when $\eta \to 0$ the system approaches total order as $\langle \psi \rangle \simeq 1 - 1.64 \, \eta^2 N$, while $\langle \sigma_{\theta}^2 \rangle$ vanishes as $\langle \sigma_{\theta}^2 \rangle \sim \eta N^{1/2}$.  This result also shows that complete order $\langle \psi \rangle = 1$ is only achieved for perfect copying $\eta=0$. 

As mentioned in section~\ref{S2}, the $2$-state case of our MSVM is equivalent to the noisy voter model studied in \cite{Considine-1989}.  This suggests that it might be possible to map the MSVM with $S>2$ states to the $2$-state noisy voter model studied in \cite{Considine-1989,Peralta-2018} by finding appropriate copying and noise rates that depend on $\eta$, and then use known results on these studied models to derive the scaling relations obtained in this article.  We have become aware of a recent unpublished article \cite{Herrerias-2019} that investigates multi-state noisy voter models where imitation and mutation (noise) events occur at respective rates $r_{ji}$ and $\epsilon_{ji}$ that may depend on states $j$ and $i$.  It seems that our model corresponds to an  homogeneous $r_{ji}$ and a very particular choice of $\epsilon_{ji}$ that depends on the fraction of particles $x_{i+1}$ and $x_{i-1}$.  This particular case is not explored by the authors who rather focus their study on the multistability properties of the system considering rates that are independent on the particles' fractions $\{x\}$.
We also need to mention that the version of the MSVM with continuous states studied in section~\ref{continuum} is related to a family of processes with $N$ branching particles, initially proposed in \cite{Brunet-1997} to investigate selection mechanisms in biological systems and later extended to a continuous time version known as ``N-branching Brownian motions'' recently explored in \cite{Maillard-2016,DeMasi-2017-1,DeMasi-2017-2}.  However, these models introduce a type of asymmetric copying process that gives rise to a traveling wave of particles that moves to the right, which is absent in our model due to the symmetry of interactions.

It is also worth mentioning some possible implications that the studied model could have on some related problems.  Within the context of flocking dynamics, the appearance of complete disorder for $\eta>0$ in the thermodynamic limit suggests an order-disorder transition at zero noise $\eta=0$, something unseen in related Vicsek-type models where a transition occurs at a finite critical value $\eta_c>0$.  Within the context of population genetics, the addition of imperfection in the process of gene replication would lead to a population characterized by a diversity of gene types that would increase with the population size.  Given that the results in this article are of mean-field type (all to all interactions), it should be worthwhile to study the imperfect MSVM in two-dimensional systems to investigate the effects of spatial interactions on the ordering dynamics.  Finally, it might be interesting to study the imperfect copying mechanism on a constrained version of the MSVM \cite{Vazquez-2003,Lanchier-2017}, where interactions are only allowed between agents whose opinion distance is smaller than a fixed threshold.  These are possible topics for further investigation.

\section*{Acknowledgments}

We acknowledge financial support from CONICET (PIP 11220150100039CO) and (PIP 0443/2014).  We also acknowledge support from Agencia Nacional de Promoci\'on Cient\'ifica y Tecnol\'ogica (PICT-2015-3628) and (PICT 2016 Nro 201-0215).

\clearpage

\appendix

\section{Power sums}
\label{power}

Below we write closed expressions for five different power sums that are useful in several calculations along the article.

\begin{eqnarray}
  \label{sum1}
  \sum_{m=0}^{M} m &=& \frac{M(M+1)}{2}, \\
  \label{sum2}
  \sum_{m=0}^{M} m^2 &=& \frac{M(M+1)(2M+1)}{6}, \\
  \label{sum3}  
  \sum_{m=0}^{M} m^3 &=& \frac{M^2 (M+1)^2 }{4}, \\
  \label{sum4}
  \sum_{m=0}^{M} m^4 &=& \frac{M(M+1)(2M+1)(3M^2+3M-1)}{30} ~~~ \mbox{and} \\
  \label{sum5}  
  \sum_{m=0}^{M} m^5 &=& \frac{M^2(M+1)^2(2M^2+2M-1)}{12}.
\end{eqnarray}

\section{Calculation of the moments $z_n$}
\label{moments}

In this section we calculate approximate expressions for the moments $z_n = \langle x_k x_j \rangle$ at the stationary state, for any $S \ge 2$ and $N \gg 1$.  For that, we derive a set of coupled equations that relate the first and second moments using the Fokker-Planck equation derived in section~\ref{fokker-planck}.  We now illustrate this procedure for the simplest case $S=2$.  From Eq.~(\ref{fpe-3}), the time evolution of the first moment $\langle x_0 \rangle$ obeys the equation
\begin{eqnarray}
 \frac{d \langle x_0 \rangle(t)}{dt} &=& \int_0^1 dx_0 \, x_0 \, \frac{\partial}{\partial t} P(x_0,t) = \frac{1}{2} \int_0^1 dx_0 \, x_0 \, \frac{\partial}{\partial x_0} \left[(2x_0-1) P(x_0,t) \right] 
  \\  &+& \frac{1}{4N} \int_0^1 dx_0 \, x_0 \, \frac{\partial^2}{\partial x_0^2} P(x_0,t), \nonumber 
\end{eqnarray}
and thus at the stationary state we have
\begin{eqnarray}
  0 = \int_0^1 dx_0 \, x_0 \, \frac{\partial}{\partial x_0} \left[(2x_0-1) P_{st}(x_0) \right] 
  + \frac{1}{2N} \int_0^1 dx_0 \, x_0 \, \frac{\partial^2}{\partial x_0^2} P_{st}(x_0),
  \label{dxdt0}
\end{eqnarray}
where $P_{st}(x_0)$ is the stationary distribution given by Eq.~(\ref{Pst}).  The above integrals can be exactly calculated using $P_{st}(x_0)$ but, instead, we can perform the integrals by parts and assume that $P_{st}(x_0)$ and its derivatives are zero at the boundaries
\begin{eqnarray}
  \label{BCs0}  
   P_{st}(x_0) |_{x_0=0} = 0, ~~ P_{st}(x_0) |_{x_0=1} = 0, \\
  \label{BCs1}
  \left. \frac{\partial}{\partial x_0} P_{st}(x_0) \right|_{x_0=0} = 0, ~~ 
  \left. \frac{\partial}{\partial x_0} P_{st}(x_0) \right|_{x_0=1} = 0.
\end{eqnarray}
One can check directly from Eq.~(\ref{Pst}) that these boundary conditions are satisfied in the $N \to \infty$ limit.  The reason is that $P_{st}(x_0)$ is a Gaussian of width $\frac{1}{2} N^{-1/2}$ centered at $x_0=1/2$.  Therefore, when $N$ is very large $P_{st}(x_0)$ quickly drops to zero outside the interval $[1/2-N^{-1/2},1/2+N^{-1/2}]$, and thus $P_{st}(x_0)$ and $\partial P_{st}(x_0)/\partial x_0$ are expected to be similar to zero at $x_0=0$ and $x_0=1$.  Then, integrating by parts Eq.~(\ref{dxdt0}) we obtain
\begin{eqnarray}
  0 &=& x_0 (2x_0-1) P_{st}(x_0) |_{0}^{1} - \int_0^1 dx_0 \, (2x_0-1) P_{st}(x_0) \nonumber \\ &+& \frac{1}{2N} \Bigg\{ x_0 \, \left. \frac{\partial P_{st}(x_0)}{\partial x_0} \right|_{0}^{1} - \int_0^1 dx_0 \, \frac{\partial P_{st}(x_0)}{\partial x_0} \Bigg\}.
  \label{dxdt1}
\end{eqnarray}
Using the boundary conditions Eqs.~(\ref{BCs0}--\ref{BCs1}) we see that only the second term of Eq.~(\ref{dxdt1}) is not zero, leading to the simple relation 
\begin{eqnarray}
  0 = - 2 \langle x_0 \rangle + 1, 
\end{eqnarray}
from where the first moments read
\begin{equation}
  \langle x_0 \rangle = \frac{1}{2} ~~~ \mbox{and} ~~~ \langle x_1 \rangle = 1 - \langle x_0 \rangle = \frac{1}{2}.
  \label{x0}
\end{equation}
Following the same approach for $d \langle x_0^2 \rangle/dt$ we obtain the relation
\begin{equation}
  0 = - 4 \langle x_0^2 \rangle + 2 \langle x_0 \rangle + \frac{1}{N}.
  \label{x02}
\end{equation}
Then, combining Eqs.~(\ref{x0}) and (\ref{x02}) we obtain the following second moments for the $S=2$ case: 
\begin{eqnarray}
  \label{z0}
  z_0 &=& \langle x_0^2 \rangle = \frac{1}{4} \left(1+\frac{1}{N} \right) ~~~ \mbox{and} \\
  \label{z1}
  z_1 &=& \langle x_0 x_1 \rangle = \langle x_0 (1-x_0) \rangle = \frac{1}{2} - z_0 = \frac{1}{4} \left(1-\frac{1}{N} \right),
\end{eqnarray}
quoted in Eq.~(\ref{zn2}) of the main text.  We can now apply the same procedure to calculate the moments $z_n$ ($0 \le n \le S-1$) for the general case scenario $S \ge 3$.  Even though the calculations are analogous to the ones for the $S=2$ case described above, the generalization is not straight forward because new types of integrals appear due the existence of crossed derivatives for $S \ge 3$.  Same as before, the idea is to write a differential equation for the time evolution of each of the moments, $\langle x_k \rangle$, $\langle x_k x_j \rangle$ and $\langle x_k^2 \rangle$, using the Fokker-Planck equation~(\ref{fpe-2}).  We illustrate here this procedure for the second moment $\langle x_l x_m \rangle$ ($l \ne m$), and leave for the interested reader the corresponding calculations for the other two moments.  From the Fokker-Planck equation~(\ref{fpe-2}) for $S \ge 3$, the time evolution of $\langle x_l x_m \rangle$ obeys the equation
\begin{eqnarray}
   \frac{d\langle x_l x_m \rangle}{dt} &=& \prod_{i=0}^{S-2} \int_0^1 dx_i \, x_l \, x_m \, \partial_t P( \{x\},t) \nonumber \\  &=& - \sum_{k=0}^{S-2} I_{lmk}(t) + \frac{1}{2} \sum_{k=0}^{S-2}  I_{lmkk}(t) + \sum_{k=0}^{S-3} \sum_{j > k}^{S-2} I_{lmkj}(t),
  \label{dxmxldt}
\end{eqnarray}
where
\begin{eqnarray}
    \label{Ilmk}
     I_{lmk}(t) &=& \prod_{i=0}^{S-2} \int_0^1 dx_i \, x_l \, x_m \, \partial_k \left[ A_k \,
      P( \{x\},t) \right], \\
    \label{Ilmkk}
     I_{lmkk}(t) &=& \prod_{i=0}^{S-2} \int_0^1 dx_i \, x_l \, x_m \, \partial_{k k}^2 \left[ B_{kk} \, P( \{x\},t) \right], ~~\mbox{and} \\
    \label{Ilmkj}
     I_{lmkj}(t) &=& \prod_{i=0}^{S-2} \int_0^1 dx_i \, x_l \, x_m \, \partial_{k j}^2 \left[ B_{kj} \, P(\{x\},t) \right].
  \end{eqnarray}
Then, at the stationary state we have 
\begin{equation}
0 = - \sum_{k=0}^{S-2} \tilde I_{lmk} + \frac{1}{2} \sum_{k=0}^{S-2}  \tilde I_{lmkk} + \sum_{k=0}^{S-3} \sum_{j > k}^{S-2} \tilde I_{lmkj},
\label{sumI}
\end{equation}
where the integrals $\tilde I_{lmk}$, $\tilde I_{lmkk}$ and $\tilde I_{lmkj}$ have the same form as those from Eqs.~(\ref{Ilmk}), (\ref{Ilmkk}) and (\ref{Ilmkj}), respectively, but integrating over the stationary distribution $P_{st}(\{x\})$ instead of $P(\{x\},t)$.  To calculate these integrals we are going to assume that $P_{st}(\{x\})$ and its first derivatives are zero at the boundaries
\begin{eqnarray}
  \label{BC0}
   P_{st}(\{x\}) |_{x_k=0} &=& 0, ~~ P_{st}(\{x\}) |_{x_k=1}=0, \\
  \label{BC1}
   \partial_j P_{st}(\{x\}) |_{x_k=0} &=& 0, ~~ 
  \partial_j P_{st}(\{x\}) |_{x_k=1} = 0, ~~ \mbox{for all $j,k=0,..,S-1$.}
\end{eqnarray} 
This is because, in analogy to the $S=2$ case, we expect for 
$S \ge 3$ a bell-shaped $P_{st}(\{x\})$ peaked at the point ($x_k=1/S \,\, \forall k$) of the $(S-1)$--dimensional space $\{x_0,..,x_{S-1} \} \in [0,1]^{S-1}$, and that the width of $P_{st}(\{x\})$ vanishes as $N^{-1/2}$ with the system size.  Therefore, if $N^{-1/2}$ is much smaller than the distance $1/S$ that separates the location of the peak and the closest boundary $x_k=0$, then $P_{st}(\{x\})$ and its first derivatives should be similar to zero at both boundaries $x_k=0$ and $x_k=1$ for all $k$.  This allows to give the rough estimation $N \gg S^2$ that relates the system size $N$ and the number of angular states $S$ for which the approximations we make in this section are valid.  In the next three subsections we calculate the integrals $\tilde I_{lmk}$, $\tilde I_{lmkk}$ and $\tilde I_{lmkj}$.

\subsection{Calculation of $\tilde I_{lmk}$}

Case $k \ne l \ne m$:
\begin{eqnarray}
  \tilde I_{lmk} = \prod_{\substack{i=0 \atop i \ne k}}^{S-2} \int_0^1 dx_i \, x_l \, x_m \int_0^1 dx_k \, \partial_k \left[ A_k P_{st} \right] = \prod_{\substack{i=0 \atop i \ne k}}^{S-2} \int_0^1 dx_i \, x_l \, x_m \Bigg\{ A_k P_{st} |_{x_k=0}^{x_k=1} \Bigg\} =0.
\end{eqnarray}
where we have used the simplified notation $P_{st}=P_{st}(\{x\})$, and the  boundary condition Eqs.~(\ref{BC0}) to set to the term inside the curly brackets to zero.

Case $k=l$:
\begin{eqnarray*}
  \tilde I_{lml} &=& \prod_{\substack{i=0 \atop i \ne l}}^{S-2} \int_0^1 dx_i \, x_m \int_0^1 dx_l \, x_l \, \partial_l \left[ A_l P_{st} \right] = \prod_{\substack{i=0 \atop i \ne l}}^{S-2} \int_0^1 dx_i \, x_m \Bigg\{ x_l \, A_l P_{st} |_{x_l=0}^{x_l=1} - \int_0^1 dx_l \, A_l P_{st}    \Bigg\} \\
   &=& - \prod_{i=0}^{S-2} \int_0^1 dx_i \, x_m \, A_l \, P_{st} = - \langle x_m  A_l \rangle.
\end{eqnarray*}
Similarly, we can show that for $k=m$ is $\tilde I_{lmm} = - \langle x_l A_m \rangle$.  Then, combining all cases we can write
\begin{eqnarray}
  \tilde I_{lmk} = - \langle x_m A_k \rangle \, \delta_{k,l} - \langle x_l A_k \rangle \, \delta_{k,m}.
  \label{Ilmks}
\end{eqnarray}

\subsection{Calculation of $\tilde I_{lmkk}$}

Case $k \ne l \ne m$:
\begin{eqnarray*}
   \tilde I_{lmkk} = \prod_{\substack{i=0 \atop i \ne k}}^{S-2} \int_0^1 dx_i \, x_l \, x_m \int_0^1 dx_k \, \partial_{kk}^2 \left[ B_{kk} P_{st} \right] = \prod_{\substack{i=0 \atop i \ne k}}^{S-2} \int_0^1 dx_i \, x_l \, x_m \Bigg\{ \partial_k \left[ B_{kk} P_{st} \right] |_{x_k=0}^{x_k=1} \Bigg\} =0.
\end{eqnarray*}
where we have used the boundary conditions Eqs.~(\ref{BC0}--\ref{BC1}) to set to the term inside the brackets to zero.

Case $k=l$:
\begin{eqnarray*}
 \tilde I_{lmll} &=& \prod_{\substack{i=0 \atop i \ne l}}^{S-2} \int_0^1 dx_i \, x_m \int_0^1 dx_l \, x_l \, \partial_{ll}^2 \left[ B_{ll} P_{st} \right] = \prod_{\substack{i=0 \atop i \ne l}}^{S-2} \int_0^1 dx_i \, x_m \Bigg\{ x_l \, \partial_l \left[ B_{ll} P_{st} \right] |_{x_l=0}^{x_l=1} \\  &-& \int_0^1 dx_l \, \partial_l \left[ B_{ll} P_{st} \right] \Bigg\} 
= - \prod_{\substack{i=0 \atop i \ne l}}^{S-2} \int_0^1 dx_i \, x_m \Bigg\{
   B_{ll} P_{st} |_{x_l=0}^{x_l=1} \Bigg\}= 0.
\end{eqnarray*}
Similarly, for $k=m$ we obtain $\tilde I_{lmmm} = 0$.  Then, 
\begin{eqnarray}
  \tilde I_{lmkk} = 0 ~~ \forall k.
  \label{Ilmkks}
\end{eqnarray}

\subsection{Calculation of $\tilde I_{lmkj}$}

Case $k \ne l \ne m \ne j$:
\begin{eqnarray*}
    \tilde I_{lmkj} = \prod_{\substack{i=0 \atop i \ne k}}^{S-2} \int_0^1 dx_i \, x_l \, x_m \int_0^1 dx_k \, \partial_{kj}^2 \left[ B_{kj} P_{st} \right] = \prod_{\substack{i=0 \atop i \ne k}}^{S-2} \int_0^1 dx_i \, x_l \, x_m \Bigg\{ \partial_j \left[ B_{kj} P_{st} \right] |_{x_k=0}^{x_k=1} \Bigg\} =0.
\end{eqnarray*}
where we have used the boundary conditions Eqs.~(\ref{BC0}--\ref{BC1}) to set to the term inside the brackets to zero.

Case $k=l, j = m \ne k$:
\begin{eqnarray*}
   \tilde I_{lmlm} &=& \prod_{\substack{i=0 \atop i \ne l}}^{S-2} \int_0^1 dx_i \, x_m \int_0^1 dx_l \, x_l \, \partial_{lm}^2 \left[ B_{lm} P_{st} \right] \\  &=& \prod_{\substack{i=0 \atop i \ne l}}^{S-2} \int_0^1 dx_i \, x_m \Bigg\{ x_l \, \partial_m \left[ B_{lm} P_{st} \right] |_{x_l=0}^{x_l=1} - \int_0^1 dx_l \, \partial_m \left[ B_{lm} P_{st} \right] \Bigg\} \\
  &=& - \prod_{\substack{i=0 \atop i \ne m}}^{S-2} \int_0^1 dx_i \int_0^1 dx_m \, x_m \, \partial_m \left[ B_{lm} P_{st} \right] \\  &=& - \prod_{\substack{i=0 \atop i \ne m}}^{S-2} \int_0^1 dx_i \Bigg\{ x_m \, B_{lm} P_{st} |_{x_m=0}^{x_m=1} - \int_0^1 dx_m \, B_{lm} P_{st} \Bigg\} = \langle B_{lm} \rangle. 
\end{eqnarray*}
Following the same type of calculations we find $\tilde I_{lmlj}=0 $ for $k=l$ and $\tilde I_{lmkm}=0$ for $j=m$. Then, 
\begin{eqnarray}
  \tilde I_{lmkj} = \langle B_{kj} \rangle \, \delta_{k,l} \, \delta_{j,m} ~~ \forall k \ne j.
  \label{Ilmkjs}
\end{eqnarray}
Finally, plugging expressions~(\ref{Ilmks}), (\ref{Ilmkks}) and (\ref{Ilmkjs}) for $\tilde I_{lmk}$, $\tilde I_{lmkk}$ and $\tilde I_{lmkj}$, respectively, into Eq.~(\ref{sumI}) we obtain the following equation for the moments:
\begin{eqnarray}
  0 &=& \sum_{k=0}^{S-2} \langle x_m A_k \rangle \, \delta_{k,l}  + \langle x_l A_k \rangle \, \delta_{k,m} + \sum_{k=0}^{S-3} \sum_{j > k}^{S-2} \langle B_{kj} \rangle \, \delta_{k,l} \, \delta_{j,m} \nonumber \\
&=& \langle x_m A_l \rangle + \langle x_l A_m \rangle + \langle B_{lm}. \rangle
\end{eqnarray}
Now that we have calculated, starting from Eq.~(\ref{dxmxldt}) for $d \langle x_m  x_l \rangle/dt$, the first equation that relates the moments, it is possible to obtain two more equations by following the same procedure for
$d \langle x_l \rangle/dt$ and for $d \langle x_l^2 \rangle/dt$ (calculations not shown).  This results in the following system of equations 
\begin{eqnarray}
  0 &=& \langle A_k \rangle, \nonumber \\
  \label{ABk-ave}  
  0 &=& 2 \langle x_k A_k \rangle + \langle B_{kk} \rangle, \\
  0 &=& \langle x_k A_j \rangle + \langle x_j  A_k \rangle + \langle B_{kj} \rangle, \nonumber 
\end{eqnarray}
where indexes $l$ and $m$ were renamed as $k$ and $j$.  Plugging expressions for $A_k$, $A_j$, $B_{kk}$ and $B_{kj}$ from Eqs.~(\ref{ABk}) into Eqs.~(\ref{ABk-ave}) leads to
\begin{eqnarray}
  \label{ABk1}
   0 &=& \langle x_{k-1} \rangle - 2 \langle x_k \rangle + \langle x_{k+1} \rangle, \\
  \label{ABk2}
   0 &=& 2 \left( \langle x_k x_{k-1} \rangle - 2 \langle x_k^2 \rangle + \langle x_k x_{k+1} \rangle \right) \nonumber \\  &+& \frac{1}{N} \left( 4 \langle x_k \rangle - 2 \langle x_k^2 \rangle + \langle x_{k-1} \rangle + \langle x_{k+1} \rangle - 2 \langle x_k x_{k-1} \rangle -  2 \langle x_k x_{k+1} \rangle \right),  \\
  \label{ABk3}
   0 &=& \langle x_k x_{j-1} \rangle - 2 \langle x_k x_j \rangle + \langle x_k x_{j+1} \rangle + \langle x_j x_{k-1} \rangle - 2 \langle x_j x_k \rangle + \langle x_j x_{k+1} \rangle \nonumber \\  &-& \frac{1}{N} \left( \langle x_k x_{j-1} \rangle + \langle x_k x_{j+1} \rangle + \langle x_j x_{k-1} \rangle + \langle x_j x_{k+1} \rangle + 2 \langle x_k x_j \rangle \right).
\end{eqnarray}
The solution to Eq.~(\ref{ABk1}) with periodic boundary conditions $\langle x_{-1} \rangle = \langle x_{S-1} \rangle$ and $\langle x_S \rangle = \langle x_0 \rangle$ that satisfies the constraint $\sum_{k=0}^{S-1} \langle x_k \rangle = 1$ is 
\begin{equation}
  \langle x_k \rangle = \frac{1}{S} ~~~ \forall k=0,..,S-1.
  \label{xk1S}
\end{equation}
If now we use the definition $z_n \equiv z_{j-k} \equiv \langle x_k x_j \rangle$ and $\langle x_k \rangle = 1/S$, we can rewrite Eq.~(\ref{ABk2}) in terms of $z_0$ and $z_1$, and Eq.~(\ref{ABk3}) in terms of $z_{n-1}$, $z_n$ and $z_{n+1}$.  For that, we need to take into account the identity $z_{-n} = \langle x_j x_k \rangle = \langle x_k x_j \rangle = z_n$.  Then, after some algebra and regrouping terms we arrive to the following system of equations for the moments
\begin{eqnarray}
  \label{zneq1}
  0 &=& z_1 - \frac{r}{2} z_0 + a, ~~~ \mbox{and} \\
  \label{zneq2}
  0 &=& z_{n-1} - r z_n + z_{n+1}, ~~~ \forall \, n=1,..,S-1 ~~~ \mbox{($S \ge 3$)},
\end{eqnarray}
where we defined
\begin{eqnarray}
  r &\equiv& \frac{2+1/N}{1-1/N}, ~~~ \mbox{and} \\
  a &\equiv& \frac{3}{2 S(N-1)}.
\end{eqnarray}
The system of Eqs.~(\ref{zneq1}--\ref{zneq2}) must also satisfy the constraint
\begin{equation}
  \sum_{n=0}^{S-1} z_n = \frac{1}{S},
  \label{constraint}
\end{equation}  
which is derived from the normalization condition $\sum_{k=0}^{S-1} x_k=1$ by multiplying both sides of this equality by $x_j$, then taking the average
$\langle \cdot \rangle$ at both sides, and setting $\langle x_j \rangle = 1/S$.
The number of independent equations in the system of Eqs.~(\ref{zneq1}--\ref{zneq2}) can be reduced by half by implementing the periodic property
\begin{equation}
  z_{S-n}=z_n ~~~ \forall n=1,..,S-1,
  \label{zn-per}
\end{equation}
obtained by direct calculation: $z_{S-n}=z_{n-S}=z_{(j-k)-S}=\langle x_S x_{j-k} \rangle =\langle x_0 x_{j-k} \rangle=z_{j-k}=z_n$.  We now show how the system of equations is reduced for the case of even $S$.  Equation~(\ref{zneq2}) for $n=S/2$ reads $z_{S/2+1}-r z_{S/2} + z_{S/2-1}$ which, after replacing $z_{S/2+1}$ by $z_{S/2-1}$ from the periodic relation Eq.~(\ref{zn-per}) becomes $2 z_{S/2-1} - r z_{S/2}=0$.  In general, one can prove that the equation for an index $n$ in the interval $S/2+1 \le n \le S-1$ becomes the same equation as that for index $S-n$.  Therefore, the system of Eqs.~(\ref{zneq1}--\ref{zneq2}) for $S > 3$ even is
\begin{eqnarray}
  \label{zneven1}
  0 &=& z_1 - \frac{r}{2} z_0 + a, \\
  \label{zneven2}
  0 &=& z_{n-1} - r z_n + z_{n+1} ~~~ \mbox{for $1 \le n \le S/2-1$} \\
  \label{zneven3}
  0 &=& 2 z_{S/2-1} - r z_{S/2}, ~~~ \mbox{and} \\
  \label{zneven4}
  0 &=& z_0 + 2 \sum_{n=1}^{S/2-1} z_n + z_{S/2} - \frac{1}{S},
\end{eqnarray}
where Eq.~(\ref{zneven4}) comes from the constraint Eq.~(\ref{constraint}). 
The same analysis applied to $S \ge 3$ odd leads to the following system
\begin{eqnarray}
  \label{znodd1}
  0 &=& z_1 - \frac{r}{2} z_0 + a, \\
  \label{znodd2}
  0 &=& z_{n-1} - r z_n + z_{n+1} ~~~ \mbox{for $1 \le n \le \frac{S-3}{2}$ ~ and} \\
  \label{znodd3}
  0 &=& (1-r) z_{\frac{S-1}{2}} + z_{\frac{S-3}{2}}, \\
  \label{znodd4}
  0 &=& z_0 + 2 \sum_{n=1}^{\frac{S-1}{2}} z_n - \frac{1}{S}.
\end{eqnarray}
Even though the set of equations for $S$ odd looks different from that of $S$ even, the solutions turn out to be the same, and thus we focus now on $S$ even.  The solution of Eqs.~(\ref{zneven1}--\ref{zneven4}) for any $S$ and $N$ is rather complicated, but because we are interested in the limit of $N \gg 1$ we look for solutions of the form $z_n = C \left( 1-\alpha_n \frac{1}{N} \right) + \mathcal O \left(1/N^2 \right)$, where $C$ is a constant and $\alpha_n$ are functions that depend on $n$ and $S$.  We note that this proposed ansatz agrees with the corresponding expressions found for the $S=2$ case [Eqs.(\ref{z0}) and (\ref{z1})], with $C=1/4$, $\alpha_0=-1/N$ and $\alpha_1=1/N$.  Then, to first order in $\epsilon \equiv 1/N \ll 1$ the solutions take the approximate form
\begin{equation}
  z_n \simeq C (1-\alpha_n \epsilon) ~~~ \mbox{for $0 \le n \le S/2$}.
  \label{zn1}
\end{equation}
Inserting the above expressions for $z_n$ into Eqs.~(\ref{zneven1}--\ref{zneven4}) and neglecting $\epsilon^2$ terms, we arrive to the following closed system of $S/2+2$ equations for $C$ and $\alpha_n$ with $S/2+2$ unknowns
\begin{eqnarray}
  \label{an1}
  0 &=& 2 S C (\alpha_0 - \alpha_1 - 3/2) + 3, \\
  \label{an2}
  0 &=& - \alpha_{n-1} + 2 \alpha_n - \alpha_{n+1} - 3 ~~~ \mbox{for $1 \le n \le S/2-1$}, \\
  \label{an3}
  0 &=& 2 \left( \alpha_{S/2} - \alpha_{S/2-1} \right) - 3, \\
  \label{an4}
  0 &=& S - \Bigg\{ \alpha_0 + 2 \sum_{n=1}^{S/2-1} \alpha_n + \alpha_{S/2} \Bigg\} \epsilon - \frac{1}{S C}.
\end{eqnarray}
To solve the system of Eqs.~(\ref{an1}--\ref{an4}) we define $\beta_n \equiv \alpha_n - \alpha_{n-1}$.  Then, from Eqs.~(\ref{an1}) and (\ref{an3}) we get $\beta_1=\frac{3}{2} \left(\frac{1}{S C} -1 \right)$ and $\beta_{S/2}=3/2$, respectively, while Eq.~(\ref{an2}) becomes $\beta_{n+1}=\beta_n-3$, whose solution is $\beta_n=\beta_1-3(n-1)$.  This last equation for $n=S/2$ leads to a simple relation between $S$ and $C$, from where we obtain $C=1/S^2$ and thus 
$\beta_1=\frac{3}{2}(S-1)$ and $\beta_n = \frac{3}{2}(S+1-2n)$.  Therefore, we get the relation $\alpha_n=\alpha_{n-1}+\frac{3}{2}(S+1-2n)$ which can be solved by simple iteration, leading to $\alpha_n=\alpha_0+\frac{3}{2}n(S-n)$.  Using this last expression for $\alpha_n$ in Eq.~(\ref{an4}) and setting $C=1/S^2 $ we arrive to a closed equation for $\alpha_0$, with solution $\alpha_0=(1-S^2)/4$.  Thus, the final expression for $\alpha_n$ becomes
\begin{equation}
  \alpha_n = \frac{1-S^2}{4} + \frac{3}{2} n (S-n).
  \label{alpha_n}
\end{equation}
Finally, using expression~(\ref{alpha_n}) for $\alpha_n$ and $C=1/S^2$ in Eq.~(\ref{zn1}) we obtain the following approximate expression for $z_n$ to first order in $1/N \ll 1$:       
\begin{equation}
  z_n \simeq \frac{1}{S^2} \left[1-\frac{1-S^2 + 6 n (S-n)}{4N} \right], ~~ \mbox{for $S \ge 3$ and $N \gg S^2$},
  \label{zn-1}
\end{equation}
quoted in Eq.~(\ref{zn3}) of the main text.  We can check that expression~(\ref{zn-1}) is a solution of the system of Eqs.~(\ref{znodd1}--\ref{znodd4}) for $S$ odd as well.

\section{Calculation of the coefficients $a_S$ and $b_S$}
\label{coefficients}

In this section we derive the expressions Eqs.~(\ref{as}) and (\ref{bs}) for the coefficients $a_S$ and $b_S$, respectively.  We start by rewriting these coefficients given by the summations in Eqs.~(\ref{as0}) and (\ref{bs0}), as the real part of complex numbers $\mathcal A_S$ and $\mathcal B_S$, respectively
\begin{eqnarray*}
  a_S = Re(\mathcal A_S) ~~~ \mbox{and} ~~~ b_S = Re(\mathcal B_S),
\end{eqnarray*}
where
\begin{eqnarray}
  \label{As}
  \mathcal A_S &=& \sum_{n=1}^{S-1} (S-n) \, r^n ~~~ \mbox{and} \\
  \label{Bs}
  \mathcal B_S &=& \sum_{n=1}^{S-1} n (S-n)^2 \, r^n,
\end{eqnarray}
with $r \equiv e^{i 2 \pi/S}$.  To perform the summations in Eqs.~(\ref{As}) and (\ref{Bs}) we first extend the upper limit to $n=S$ and the lower limit of Eq.~(\ref{Bs}) to $n=0$, then expand the terms in brackets and define
\begin{eqnarray*}
  y_m \equiv \sum_{n=0}^S n^m \, r^n.
\end{eqnarray*}
Thus, Eqs.~(\ref{As}) and (\ref{Bs}) can be written as
\begin{eqnarray}
  \label{As1}
  \mathcal A_S &=& S(y_0-1) - y_1  ~~~ \mbox{and} \\
  \label{Bs1}
  \mathcal B_S &=& S^2 y_1 + y_3 - 2 S y_2.
\end{eqnarray}
In order to find $y_m$ ($m=0,1,2,3$) we start from the well known geometric series
\begin{eqnarray*}
  y_0=\sum_{n=0}^S r^n=\frac{1-r^{S+1}}{1-r},
\end{eqnarray*}
and differentiate this formula with respect to $r$ to obtain
\begin{eqnarray*}
   y_1 &=& \sum_{n=0}^S n \, r^n = r \frac{\partial y_0}{\partial r} = \frac{r}{(1-r)^2} \Big\{ 1-
  \left[ S(1-r)+1 \right] r^S \Big\}, \\
   y_2 &=& \sum_{n=0}^S n^2 \, r^n = r \frac{\partial y_1}{\partial r} = \frac{r}{(1-r)^3} \Big\{ 1 + r - \left[ \left( S(1-r)+1 \right)^2 + r \right] r^S \Big\}, \\
   y_3 &=& \sum_{n=0}^S n^3 \, r^n = r \frac{\partial y_2}{\partial r} \nonumber \\ &=& \frac{r}{(1-r)^4} \Big\{ 1 + 4r + r^2 - \left[ S(1-r)+1 \right]^3 r^S +
  \left[ (3S-1)r - 3S -4 \right] r^{S+1} \Big\}.   
\end{eqnarray*}
These formulas can be greatly simplified by noting that $r^S=e^{i 2 \pi}=1$, and thus $r^{S+1}=r$ and $r^{S+2}=r^2$, which leads to
\begin{eqnarray*}
  y_0 &=& 1, \\
  y_1 &=& - \frac{S r}{1-r}, \\
  y_2 &=& - \frac{S r \left[ S(1-r)+2 \right]}{(1-r)^2}, \\
  y_3 &=& - \frac{S r \left[ S^2 (1-r)^2 + 3 S (1-r) + 3 (1+r) \right]}{(1-r)^3}.
\end{eqnarray*}
Replacing these expressions for $y_m$ in Eqs.~(\ref{As1}) and (\ref{Bs1}) we obtain
\begin{eqnarray*}
  \label{As2}
  \mathcal A_S &=& \frac{S r}{(1-r)}  ~~~ \mbox{and} \\
  \label{Bs2}
  \mathcal B_S &=& \frac{S r \left[ S(1-r) - 3(1+r) \right]}{(1-r)^3}.
\end{eqnarray*}
Substituting in the above expressions $r$ by $e^{i 2 \pi/S}$ and using for convenience the identities $1-r=- 2 \, i \, e^{i \pi/S} \sin(\pi/S)$ and $1+r= 2 \, e^{i \pi/S} \cos(\pi/S)$ we finally arrive to 
\begin{eqnarray*}
  \label{As3}
  \mathcal A_S &=& - \frac{S}{2} + i \, \frac{S}{2 \tan(\pi/S)}  ~~~ \mbox{and} \\
  \label{Bs3}
  \mathcal B_S &=& - \frac{S^2}{4 \sin^2(\pi/S)} + i \, \frac{3 \cos(\pi/S)}{4 \sin^3(\pi/S)},
\end{eqnarray*}
whose real parts correspond to Eqs.~(\ref{as}) and (\ref{bs}), respectively, of the main text.

\bibliographystyle{apsrev}

\bibliography{references}

\end{document}